\documentclass[final,5p,times,twocolumn]{elsarticle}
\usepackage[utf8]{inputenc}
\usepackage{hyperref}
\usepackage{bm}
\usepackage{nicefrac}
\usepackage{amsmath}
\usepackage{caption}
\usepackage[capitalise]{cleveref}
\usepackage[ruled,vlined]{algorithm2e}
\usepackage{booktabs}
\usepackage{siunitx,booktabs,subcaption}

\usepackage{lipsum} % Just for sample text
\usepackage{array} % For table wrapping
\usepackage{makecell} % For line breaks in table cells

\newcommand{\authbreak}{\newline} % 著者名の改行用コマンド

\usepackage{tikz}

\hypersetup{colorlinks=true,linkcolor=blue,anchorcolor=red,citecolor=blue,filecolor=red,urlcolor=red,pdfauthor=author}

\newcommand{\abbreviations}[1]{%
  \nonumnote{\textit{Abbreviations:\enspace}#1}}

\begin{document}

\begin{frontmatter}

\title{Can Physician Judgment Enhance Model Trustworthiness?\\A Case Study on Predicting Pathological Lymph Nodes in Rectal Cancer}

\author[a,b]{Kazuma Kobayashi\corref{correspondingauthor}}
\cortext[correspondingauthor]{Corresponding to: Kazuma Kobayashi, M.D., D.Sc. {\it Postal address:} Division of Medical AI Research and Development, National Cancer Center Research Institute, 5-1-1 Tsukiji Chuo-ku, Tokyo, 104-0045, Japan; {\it Email address:} {\texttt kazumkob@ncc.go.jp}}
\ead{kazumkob@ncc.go.jp}

\author[c]{Yasuyuki Takamizawa}
\ead{ytakamiz@ncc.go.jp}

\author[d]{Mototaka Miyake}
\ead{mmiyake@ncc.go.jp}

\author[c]{Sono Ito}
\ead{sitosrg1@tmd.ac.jp}

\author[e,f]{Lin Gu}
\ead{lin.gu@riken.jp}

\author[g]{Tatsuya Nakatsuka}
\ead{8123526@ed.tus.ac.jp}

\author[h]{Yu Akagi}
\ead{yu.akagi1115@gmail.com}

\author[f,e]{\authbreak Tatsuya Harada}
\ead{harada@mi.t.u-tokyo.ac.jp}

\author[c]{Yukihide Kanemitsu}
\ead{ykanemit@ncc.go.jp}

\author[a,b]{Ryuji Hamamoto}
\ead{rhamamot@ncc.go.jp}

\address[a]{Division of Medical AI Research and Development, National Cancer Center Research Institute, \\5-1-1 Tsukiji, Chuo-ku, Tokyo 104-0045, Japan}
\address[b]{Cancer Translational Research Team, RIKEN Center for Advanced Intelligence Project, \\1-4-1 Nihonbashi, Chuo-ku, Tokyo 103-0027, Japan}
\address[c]{Department of Colorectal Surgery, National Cancer Center Hospital, \\5-1-1 Tsukiji, Chuo-ku, Tokyo 104-0045, Japan}
\address[d]{Department of Diagnostic Radiology, National Cancer Center Hospital, \\5-1-1 Tsukiji, Chuo-ku, Tokyo 104-0045, Japan}
\address[e]{Machine Intelligence for Medical Engineering Team, RIKEN Center for Advanced Intelligence Project, \\1-4-1 Nihonbashi, Chuo-ku, Tokyo 103-0027, Japan}
\address[f]{Research Center for Advanced Science and Technology, The University of Tokyo, \\4-6-1 Komaba, Meguro-ku, Tokyo 153-8904, Japan}
\address[g]{Department of Applied Electronics, Graduate School of Advanced Engineering, Tokyo University of Science, \\6-3-1 Niijuku, Katsushika-ku, Tokyo 125-8585, Japan}
\address[h]{Department of Biomedical Informatics, Graduate School of Medicine, The University of Tokyo, \\7-3-1 Hongo, Bunkyo-ku, Tokyo 113-8655, Japan}

\date{December 2023}

\begin{abstract}
Explainability is key to enhancing artificial intelligence's trustworthiness in medicine. However, several issues remain concerning the actual benefit of explainable models for clinical decision-making. Firstly, there is a lack of consensus on an evaluation framework for quantitatively assessing the practical benefits that effective explainability should provide to practitioners. Secondly, physician-centered evaluations of explainability are limited. Thirdly, the utility of built-in attention mechanisms in transformer-based models as an explainability technique is unclear. We hypothesize that superior attention maps should align with the information that physicians focus on, potentially reducing prediction uncertainty and increasing model reliability. We employed a multimodal transformer to predict lymph node metastasis in rectal cancer using clinical data and magnetic resonance imaging, exploring how well attention maps, visualized through a state-of-the-art technique, can achieve agreement with physician understanding. We estimated the model's uncertainty using meta-level information like prediction probability variance and quantified agreement. Our assessment of whether this agreement reduces uncertainty found no significant effect. In conclusion, this case study did not confirm the anticipated benefit of attention maps in enhancing model reliability. Superficial explanations could do more harm than good by misleading physicians into relying on uncertain predictions, suggesting that the current state of attention mechanisms in explainability should not be overestimated. Identifying explainability mechanisms truly beneficial for clinical decision-making remains essential.
\end{abstract}


\abbreviations{3D, three-dimensional; AI, artificial intelligence; AUC-ROC, area under the receiver operating characteristic curve; BCE, binary cross-entropy; CNN, convolutional neural network; FT, feature tokenizer; Grad-CAM, gradient-weighted class activation mapping; LN, layer normalization; LPLN, lateral pelvic lymph node; MHSA, multi-head self attention; MLN, mesorectal lymph node; MLP, multilayer perceptron; MRI, magnetic resonance imaging; ViT, vision transformer}

\end{frontmatter}

\section{Introduction}\label{sec:introduction}

\begin{figure}[t!]
  \centering
  \includegraphics[width=\columnwidth]{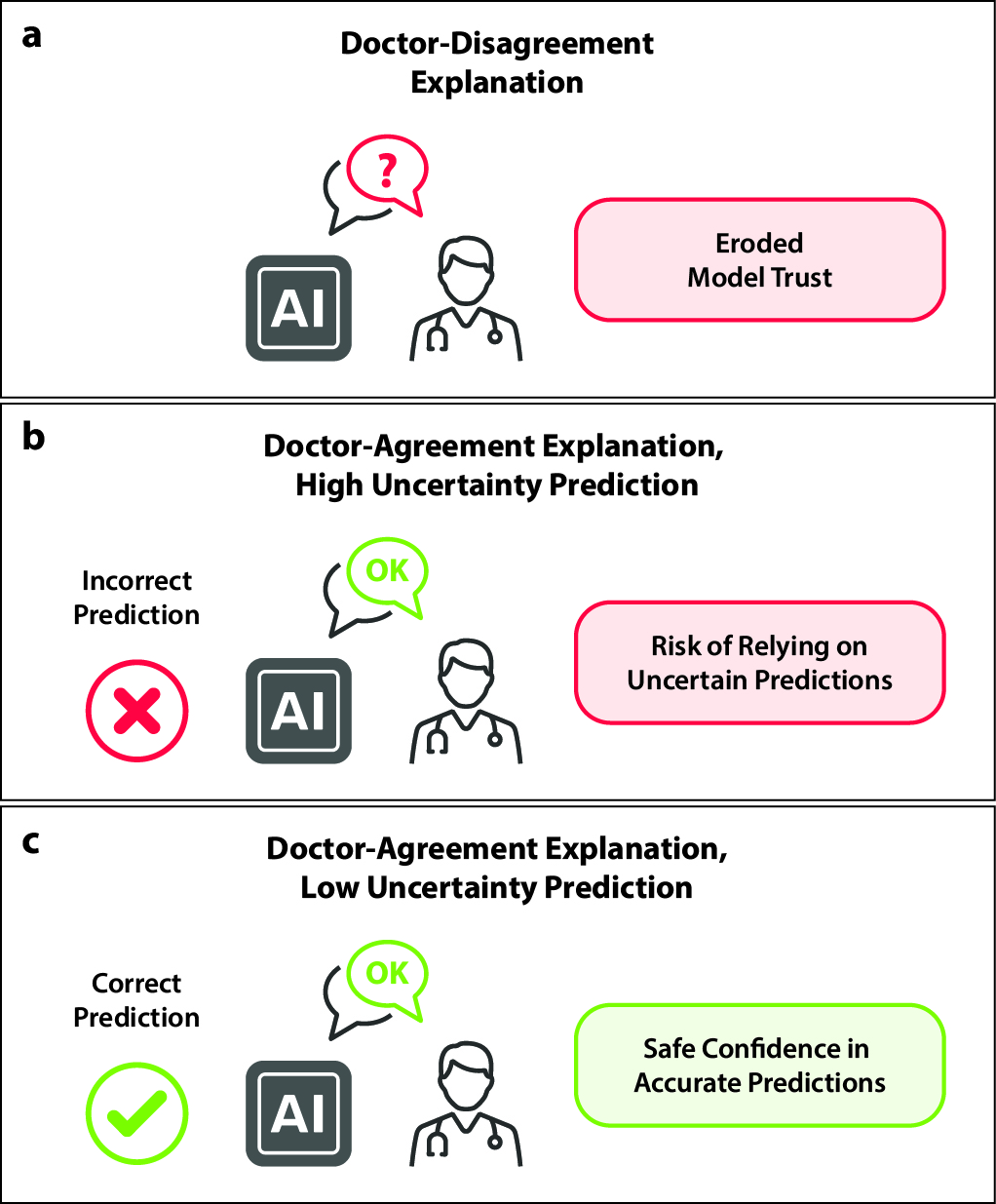}
  \caption{\textbf{Agreement between the model explanation and the physician's understanding, considered based on aspects of safety and reliability.} \textbf{(a)} If the model explanation does not align with a physician's understanding, trust in the model may be compromised, regardless of its accuracy. \textbf{(b)} Even when there is agreement between the model explanation and the physician's understanding, the reliability of the model's prediction remains questionable if this consensus does not reduce uncertainty estimation. This underscores the risks associated with relying on uncertain predictions. \textbf{(c)} Trust in the model prediction is considered safe only when the explanation aligns with the physician's understanding and effectively reduces uncertainty estimation.}
  \label{fig:framework}
\end{figure}

\begin{figure*}[t!]
  \centering
  \includegraphics[width=\textwidth]{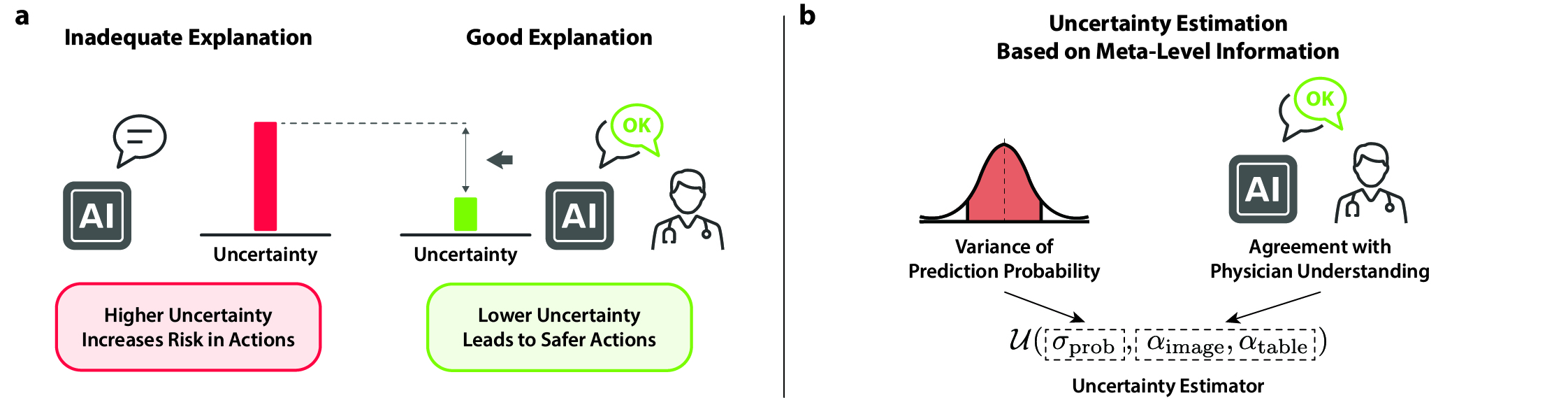}
  \caption{\textbf{Can agreement between model explanations and a physician's understanding reduce overall uncertainty?} \textbf{(a)} We hypothesize that a good explanation in agreement with a physician's understanding should also provide additional information to reduce the overall uncertainty of the model prediction, thereby leading to safer actions. In contrast, if the model explanation cannot provide such additional information, the agreement with physicians could not have any beneficial impact from a practical point of view. \textbf{(b)} To quantitatively investigate this hypothesis, we trained a machine learning-based uncertainty estimator that utilizes the variance of prediction probability, with or without additional information regarding agreement with physician understanding. We then investigated how the estimated uncertainty was affected by the additional information to assess the beneficial effect of model explanations in clinical decision-making.}
  \label{fig:question}
\end{figure*}

Explainability is one of the core essentials for achieving the trustworthiness of artificial intelligence (AI) \citep{Kundu2021}. There are numerous motivations for seeking explainability in AI. These include justifying the decision-making process, controlling system behavior, improving systems through human-AI interaction, and discovering new insights \citep{Adadi2018}. Such motivation is also vital for its successful implementation in clinical practice \citep{Cutillo2020}. In medicine, simply demonstrating accuracy is not sufficient; understanding disease mechanisms and clearly explaining decisions to patients are also essential for physicians \citep{Khullar2022, Kundu2023}. Therefore, there is a growing expectation for, and technological advancements in, explainable AI \citep{Markus2021}. These advancements aim to mitigate the so-called ``black-box'' nature of deep learning models, transforming originally opaque and inscrutable predictions into transparent and explicable ones \citep{Ghassemi2021}.

On the other hand, there remains considerable controversy over the actual benefits of explainable AI. For example, \cite{Rudin2019} argues that explanations are often unreliable and misleading. This is because techniques for explainability sometimes provide explanations that are not faithful to the computations of the original model, and either do not make sense or do not provide sufficient details for understanding the model's behavior. 

Nevertheless, it would be premature to apply these conclusions directly to the medical field. Indeed, there is a lack of frameworks for the evaluation of explainability by physicians. The actual performance of explainability has rarely been tested from a physician's perspective in medicine \citep{Ghassemi2021}. Besides, the most controversial evidence comes from saliency maps of convolutional neural networks (CNNs) \citep{Groen2022}, but the usefulness of built-in attention mechanisms in recently introduced transformers remains unclear. Based on these considerations, this study aims to address three fundamental challenges as follows: (1) the absence of an evaluation framework for assessing the practical benefits of good explainability; (2) the lack of evaluation of explainability by expert physicians; (3) limited evidence regarding the utility of the attention mechanism in recently introduced transformer-based architectures. % English proofed. 

\subsection{Evaluation Framework for Good Explanation}\label{sec:eval_framework}

There is no consensus framework for quantitatively assessing the practical benefits that good explainability should offer practitioners in clinical decision-making \citep{Zhou2021}. Recently, \cite{Kundu2023} proposed an intuitive formulation inspired by the Turing test: an AI model can be considered as trustworthy as a medical expert if it provides explanations that are consistent and well-aligned with human understanding. We interpret this formulation to mean that when there is \textbf{agreement} between model explainability and physician understanding, it enables us to confidently utilize the model's predictions in actual clinical decision-making. This is because such a model presumably captures meaningful associations in data in a manner similar to a medical expert. Conversely, when such agreement is not achieved, it becomes challenging for practitioners to fully trust the model's predictions, as the model may be learning superficial biases in the dataset that diverge from clinically meaningful relationships (see \textbf{\cref{fig:framework}a}).

Nevertheless, this formulation might not be sufficient for the safe evaluation of explainability. As exemplified by the hallucination phenomenon in recent large language models \citep{Huang2023}, explanations generated by AI systems can sometimes be superficially accurate yet fail to be deeply rooted in factual associations. Therefore, relying solely on agreement between model explainability and physician understanding to evaluate explanations carries the risk of mistakenly endorsing hallucinated explanations. As shown in \textbf{\cref{fig:framework}b}, this could be highly detrimental for practitioners responsible for clinical decision-making \citep{Lakkaraju2020}.

Then, what additional elements should be incorporated into the evaluation framework? In practice, physicians may seek explainability in a model primarily to estimate its reliability, with the assumption that a model's explanation aligning well with their clinical understanding would correlate with the accuracy of its predictions (see \textbf{\cref{fig:framework}c}). From a technical perspective, such reliability can be quantified as model uncertainty, providing a means to estimate the confidence of the prediction \citep{Loftus2022}. Thus, model explainability should be linked to uncertainty estimation through its agreement with physician understanding. Based on these considerations, we propose the following criteria for our evaluation framework (see \textbf{\cref{fig:question}a}):
\begin{itemize}
    \item When there is agreement between model explainability and physician understanding, and this agreement successfully provides additional information to reduce model uncertainty, the explanation is considered good.
    \item When agreement is not achieved, or additional information to reduce model uncertainty is not provided even after achieving agreement, the explanation is considered inadequate.
\end{itemize}

In summary, our evaluation framework expects a good explanation to contribute to reducing model uncertainty, thereby enhancing the reliability of AI-aided clinical decision-making. As shown in \textbf{\cref{fig:question}b}, the advantage of this framework is that the effectiveness of an explanation can be demonstrated by the degree to which it reduces model uncertainty, as evidenced by the additional information provided through its agreement with physician understanding.

\subsection{Evaluating Transformer Explainability by Physicians: A Case Study}\label{sec:introduction_case_study}

Herein, we have established a case study to evaluate the actual benefit of visualized attention maps from transformer-based architectures in clinical decision-making.

The task of predicting pathological lymph node metastasis from preoperative clinical information and magnetic resonance imaging (MRI) of rectal cancer is considered. The probability of pathological lymph node metastasis is a critical piece of information for deciding on surgical methods and postoperative adjuvant therapy \citep{Errit1998, Zaborowski2021, Malla2023}, which contributes to more precise and safer treatments. However, accurately predicting pathological lymph node metastasis before surgery can be challenging. For instance, small metastases, such as occult metastases \citep{Guo2012}, may not be detectable on imaging, leading to false negatives. Conversely, lymph nodes can also enlarge due to various benign causes, such as inflammation, apart from metastasis \citep{Stijns2021}. Such fundamental challenges necessitate the combination of model uncertainty and explainability with actual physician judgments, especially if the model is to be implemented in clinical decision-making.

A multimodal transformer-based architecture is used for this case study. Transformers, a deep neural network architecture based on attention mechanisms, are adept at handling long-term dependencies in context and were first introduced for natural language processing \citep{Vaswani2017}. Following the introduction of several transformer-based language models \citep{Devlin2019, Radford2018}, there has been an expansion into multimodal fusion tasks involving images and language \citep{Xu2023}. Particularly noteworthy is the trend of processing images alone with transformers, significantly influenced by Vision Transformer (ViT), which divides the image into equal patches and treats them as tokens \citep{Alexey2021}. Due to its flexible and unified architecture, the transformer-based architecture is also playing an increasingly important role in healthcare AI systems, where integrated decision-making with multimodal data is essential \citep{Zhou2023}.

The overview of our case study is as follows: We trained a multimodal transformer-based architecture to predict the likelihood of postoperative pelvic lymph node metastasis of rectal cancer using tabular clinical data combined with MRI inputs. Subsequently, we acquired two types of \textbf{meta-level information} to estimate the model's uncertainty for each prediction: the variance of prediction probability and the quantified agreement between model explainability and physician understanding. The variance in prediction probability, which is usually obtained for quantifying model uncertainty \citep{Loftus2022}, was estimated using a \emph{test-time data augmentation} approach \citep{Wang2019}. For the technique of model explainability, we utilized \emph{Transformer Explainability} \citep{Chefer2021}, a state-of-the-art method for visualizing class-specific attention maps that incorporate gradient information. This visualization was applied to both tabular clinical data and MRI inputs. Expert physicians then evaluated the visualized attention maps as the model explanation based on the agreement with the actual referenced information, leading to the creation of the \textbf{agreement score}.

We then performed an evaluation to determine if additional information, based on the agreement between the model explanation and physician understanding, could reduce the overall uncertainty. Based on the idea illustrated in \textbf{\cref{fig:question}b}, we employed a machine learning-based model. This model, termed the \textbf{uncertainty estimator}, estimates the \textbf{uncertainty score}, which is a normalized measure of model uncertainty derived from the probability of correctness for each prediction. The uncertainty score is computed based on the variance of prediction probability, either including or excluding the quantified agreement as additional information. Hence, the difference depending on the additional information can reflect the anticipated benefit from the model explainability. However, we found that the additional information had no significant impact on the estimation of the uncertainty score. 

Consequently, the model's reliability cannot be guaranteed by its explanation, which might be misleading for practitioners relying on its apparent understandability (see \textbf{\cref{fig:framework}b}). This study sheds light on the current limitations of explainability techniques in healthcare and underscores the need for their development to enhance model trustworthiness from more human-centered perspectives.

\subsection{Contributions}\label{sec:contributions}
The main contributions are as follows:
\begin{itemize}
\item We propose an evaluation framework for assessing the effectiveness of explainability in clinical decision-making, emphasizing the importance of agreement between model explanation and physician understanding. This agreement is expected to provide additional information that reduce uncertainty in the model's predictions, thereby enhancing its reliability.
\item By introducing the agreement score and uncertainty score, we quantitatively assessed the impact of alignment between the model's explanation and physician understanding on model reliability. This evaluation was conducted in a case study using a multimodal transformer trained to predict pathological lymph node metastasis in rectal cancer based on preoperative information.
\item We found that even explanations with higher agreement with physician understanding do not necessarily contribute additional information to reduce model uncertainty. This observation highlights significant room for improvement in explainability techniques, making them more beneficial for clinical decision-making.
\end{itemize}

\section{Preliminaries}\label{sec:preliminaries}

Here, we introduce prior works relevant to our research and their implications for our case study.

\subsection{Inherently Interpretable Models and Post Hoc Explanations}\label{sec:post_hoc_explanations}

Technical approaches to model explainability are often categorized into two types: \emph{inherently interpretable models} and \emph{post hoc explanations}. Inherently interpretable models incorporate explanation mechanisms within their internal architecture, designed to be self-explanatory and capable of producing human-understandable representations from internal model features \citep{Patr2023}. Examples of such models include decision trees, prototype-based models \citep{Kim2015}, and generalized additive models \citep{Caruana2015}. However, given the high predictive performance of complex models such as deep neural networks, post hoc explanations have recently become the mainstay for understanding a model's behavior. Techniques for post hoc explanations differ in how they handle the complex model (e.g., black box vs. accessible internal mechanism), the scope of interpretability (e.g., global vs. local), the search technique (e.g., perturbation-based vs. gradient-based), and the modalities of explanation (e.g., explanation by feature attribution, text, examples, concepts) \citep{Krishna2022, Patr2023}. 

Our case study employs a method that combines the visualization of the attention map from the transformer-based architecture with gradient information \citep{Chefer2021}. We test whether the additional information, derived from the agreement between model explanations and physician understanding, affects model uncertainty. It is important to note that the uncertainty quantification is rooted in a population-based statistic, rather than an indicator for a binary decision on whether or not to believe each prediction. Therefore, our focus is on investigating global interpretability for black-box models by using feature attribution as a modality of explanation, in conjunction with a gradient-based search technique.

\subsection{Evaluation of Post Hoc Explanations}\label{sec:eval_post_hoc_explanations}

According to \cite{Markus2021}, explainability is an overarching concept that encompasses various properties and requires several notions of explanation quality. These include interpretability (whether the explanation is understandable to humans) and fidelity (whether the explanation accurately describes the internal mechanism of the model). Simultaneously, various metrics have been proposed to evaluate these different aspects of explainability quality, but most of them have been designed for specific contexts, resulting in a lack of generality \citep{Gilpin2019, Zhou2021}. As a more serious issue, many of these metrics are inconsistent with each other, depending on the specific explainable methods used, and researchers have shown that they sometimes lead to harmful misconceptions for humans \citep{Zana2020, Krishna2022}.

Furthermore, there is still no consensus on the framework for evaluating the quality of explainability. Given the inherently subjective nature of the concept of model explainability, \emph{human-centered evaluations} are essential. These differ from \emph{functionality-grounded evaluations}, which rely on proxies based on a formal definition \citep{Doshivelez2017}. However, most previous studies have favored such proxy measures, primarily because recruiting expert physicians is quite costly. Consequently, the evidence based on expert physicians remains quite limited \citep{Ghassemi2021}.

Indeed, the subjectivity brought about by human evaluations can be both an advantage and a disadvantage \citep{Zana2020}. Advantages include the real-world relevance based on actual physician understanding, while disadvantages include bias and scalability issues. However, medical knowledge about clinical causal relationships, as understood by expert physicians, is deemed to be able to help models make correct inferences, warranting further investigation. Hence, we conducted evaluations of the model's explainability by several experts, centering on the agreement between model explainability and physician understanding.

\subsection{Impact of Explainability on Clinical Decision-Making}\label{sec:impact_on_clinical_decision_making}

Previous research investigating the impact of model explainability on clinical decision-making encompasses several distinct perspectives.

\vspace{2mm}
\noindent \textbf{Attributes of Explainability:}
\cite{Tonekaboni2019} extracts the desired attributes of model explainability through interviews with clinicians, which include domain representativeness, actionability, and consistency. Particularly, many clinicians believed that a correct explanation should correspond with more accurate prediction performance, considered as a prerequisite for actionability.

\vspace{2mm}
\noindent \textbf{Influence of Explanations on Physicians' Acceptance:}
This perspective investigates how explanations influence physicians' acceptance of the model's predictions. \cite{Dong2023} highlights the role of real-time explanations in enhancing physicians' trust and acceptance, while \cite{Gaube2023} suggest that non-experts may benefit more from these explanations than experts in enhancing their acceptance of the model's predictions.

\vspace{2mm}
\noindent \textbf{Correctness of Explainability from a Clinical Viewpoint:}
\cite{Tsiknakis2020} investigates the validity of saliency maps visualized by gradient-weighted class activation mapping (Grad-CAM, \citep{Selvaraju2017}) in a CNN performing COVID-19 screening on chest X-rays. Two diagnostic radiologists evaluated these saliency maps, revealing that, in some cases, the model focused on areas other than the anatomically valid locations. However, this study did not systematically examine the relationship between the validity of the explanations and the accuracy of the model's predictions.

\vspace{2mm}
From each of these perspectives, good explanations are anticipated to enhance the model's trustworthiness, thereby fostering a more accepting attitude. Nonetheless, the correlation between the quality of explanations and the actual predictive accuracy of the model, which is essential for improved acceptance, has not been verified. In fact, \cite{Ehrmann2023} propose that encouraging users to take action based on model predictions should be grounded in reducing model uncertainty. Therefore, it is important to quantitatively verify whether the appropriateness of explainability, as judged by physicians, can serve as additional information to reduce model uncertainty.

\subsection{Explainability in Transformers}\label{sec:explainability_transformers}

While most research on model explainability still predominantly focuses on saliency maps in CNN-based architectures \citep{Groen2022}, the exploration of explainability in transformers has become increasingly important \citep{Patr2023}. Indeed, the built-in attention mechanism in transformers is a critical method for elucidating how models make predictions, as it visualizes which variables or pixels in the input are focused on. However, as noted in several studies \citep{Jain2019, Wiegreffe2019}, selecting the appropriate layers for attention visualization can be challenging. Besides, many of the visualization methods for the attention maps have not been class-specific, thereby returning the same visualization regardless of the interested class. In our case study, we explored the state-of-the-art method known as Transformer Explainability \citep{Chefer2021}, which utilizes a relevancy propagation rule applicable to both positive and negative attributes and is class-specific by design.

\subsection{Uncertainty Quantification in Transformers}\label{sec:uncertainty_transformers}

\begin{figure*}[t!]
  \centering
  \includegraphics[width=0.75\textwidth]{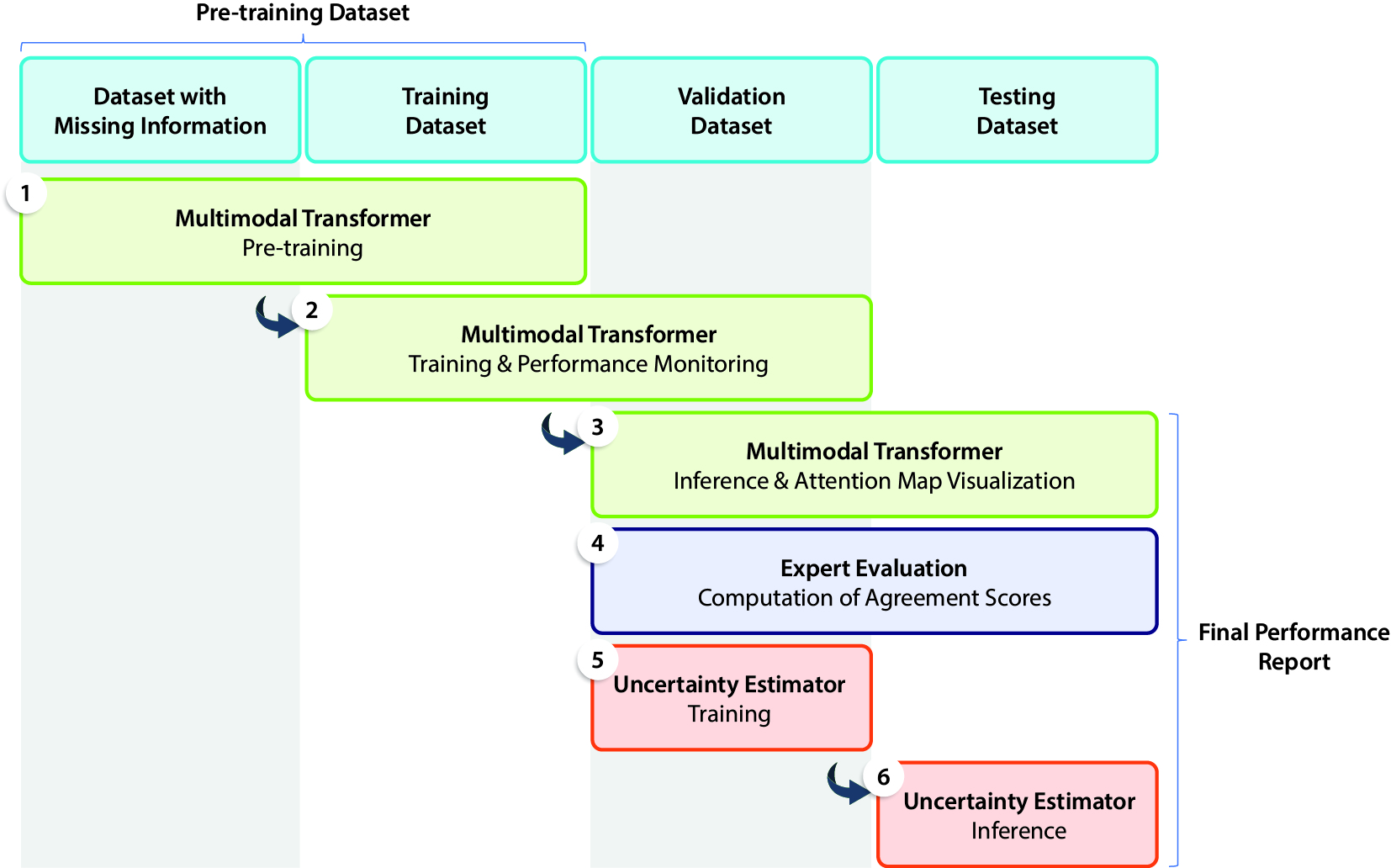}
  \caption{\textbf{The six steps of dataset usage in this case study.} \textbf{(1)} The multimodal transformer was pre-trained using a pre-training dataset, which comprises both the dataset with missing information and the training dataset. \textbf{(2)} The multimodal transformer was then trained with the training dataset, and its performance was monitored using the validation dataset. \textbf{(3)} Inference by the trained multimodal transformer was performed on both the validation and testing datasets, yielding inference results, such as prediction probabilities, their variance, and visualized attention maps. \textbf{(4)} Expert physicians evaluated the visualized attention maps on both the validation and testing datasets to compute agreement scores. \textbf{(5)} The uncertainty estimator was trained using meta-level information, namely, the variance in prediction probabilities with or without the quantified agreement scores, derived from the validation dataset. \textbf{(6)} A final performance evaluation was conducted by combining the inference results of the multimodal transformer and the uncertainty estimator on the testing dataset.}
  \label{fig:dataset}
\end{figure*}

Uncertainty estimation is essential for producing confidence along with the prediction of deep neural networks, which can also play an important role in enhancing the model's trustworthiness. In general, technological approaches to estimate uncertainty include \citep{Zou2023}: the \emph{deterministic method}, which provides a deterministic estimate of uncertainty; \emph{bayesian neural networks}, which explicitly compute uncertainty by introducing random variables into the model parameters; the \emph{ensemble method}, which aggregates predictions from multiple models and estimates their variance; and \emph{test-time data augmentation}, which involves augmenting the test data with different perturbations and evaluating its variance.

Despite such advancements, uncertainty quantification in transformers represents another significant hurdle. Transformers are known for their tendency to overfit, which poses a challenge in accurately quantifying uncertainty \citep{Ma2022}. Although uncertainty quantification remains a heavily researched field aimed at enhancing the reliability of model predictions across various architectures \citep{Gal2016, Gal2016_RNN}, our understanding of these techniques in transformers is still limited \citep{Sankararaman2022}. This issue becomes particularly pronounced with limited datasets, a common scenario in many clinical tasks.

Based on the test-time data augmentation approach, we investigated several methods known to enhance the generalization capability of neural networks in small datasets, including \emph{cosine loss} \citep{Barz2020}. Our exploration enabled us to identify the learning conditions under which multimodal transformers can output model uncertainty, which statistically correlated well with the correctness of predictions (see \textbf{\cref{sec:performance_base_model}}). This discovery is useful as a baseline for determining whether additional benefits can be gained from uncertainty estimation when there is an agreement between model explainability and physician understanding.

\section{Experiments}\label{sec:experiments}

In this section, we primarily describe the training and inference methods for the multimodal transformer, which is expected to predict pathological lymph node metastases from tabular clinical data and MRI inputs. We also present the techniques for calculating the variance of prediction probability and visualizing attention maps.

\subsection{Dataset Preparation}\label{sec:dataset}

The study, data use, and data protection procedures were approved by the Ethics Committee of the National Cancer Center, Tokyo, Japan (protocol number 2016-496). We collected data from 900 rectal cancer patients at a single institution, each with an MRI scan and a bounding box that encompasses the site of the primary tumor. Expert physicians annotated the bounding boxes on the T2-weighted image sequences of the MRIs. The majority of the patient data included tabular clinical data and an MRI scan, both obtained during the preoperative evaluation for each patient. As shown in \textbf{\cref{tab:preoperative_clinical_data}}, the tabular clinical data contain 18 variables, encompassing demographic information, tumor markers, and other tumor assessments obtained during the preoperative evaluation, including endoscopy and radiological studies. In clinical practice, all targeted patients underwent surgery, which included total mesorectal excision, with or without lateral lymph node dissection. Postoperatively, the pathological findings, including the state of pelvic lymph node metastasis (encompassing the left and right lateral pelvic lymph node (LPLN) and mesorectal lymph node (MLN)), were revealed. For the patient who underwent surgery without lateral lymph node dissection, the metastatic statuses of the left and right LPLNs were considered negative.

As shown in \textbf{\cref{fig:dataset}}, the dataset is utilized based on the following split: First, the 900 cases were categorized based on the completeness of their clinical information. This categorization resulted in 79 cases with some missing values in their clinical data and 821 cases with complete clinical information. Then, the dataset of 821 cases was randomly divided into a \emph{training dataset} of 661 patients, a \emph{validation dataset} of 60 patients, and a \emph{testing dataset} of 100 patients. Additionally, the 79 cases with incomplete clinical information were added to the training dataset to create a \emph{pre-training dataset} with 740 patients. Note that the pre-training dataset fully equips only MRIs with bounding boxes encompassing the primary tumor. In contrast, other datasets (i.e., the training, validation, and testing datasets) are complete with preoperative clinical information, MRIs annotated with bounding boxes, and postoperative pathological information. See \textbf{\cref{tab:detail_dataset}} for the detailed characteristics of patients in the training, validation, and testing datasets.

\begin{figure*}[t!]
  \centering
  \includegraphics[width=\textwidth]{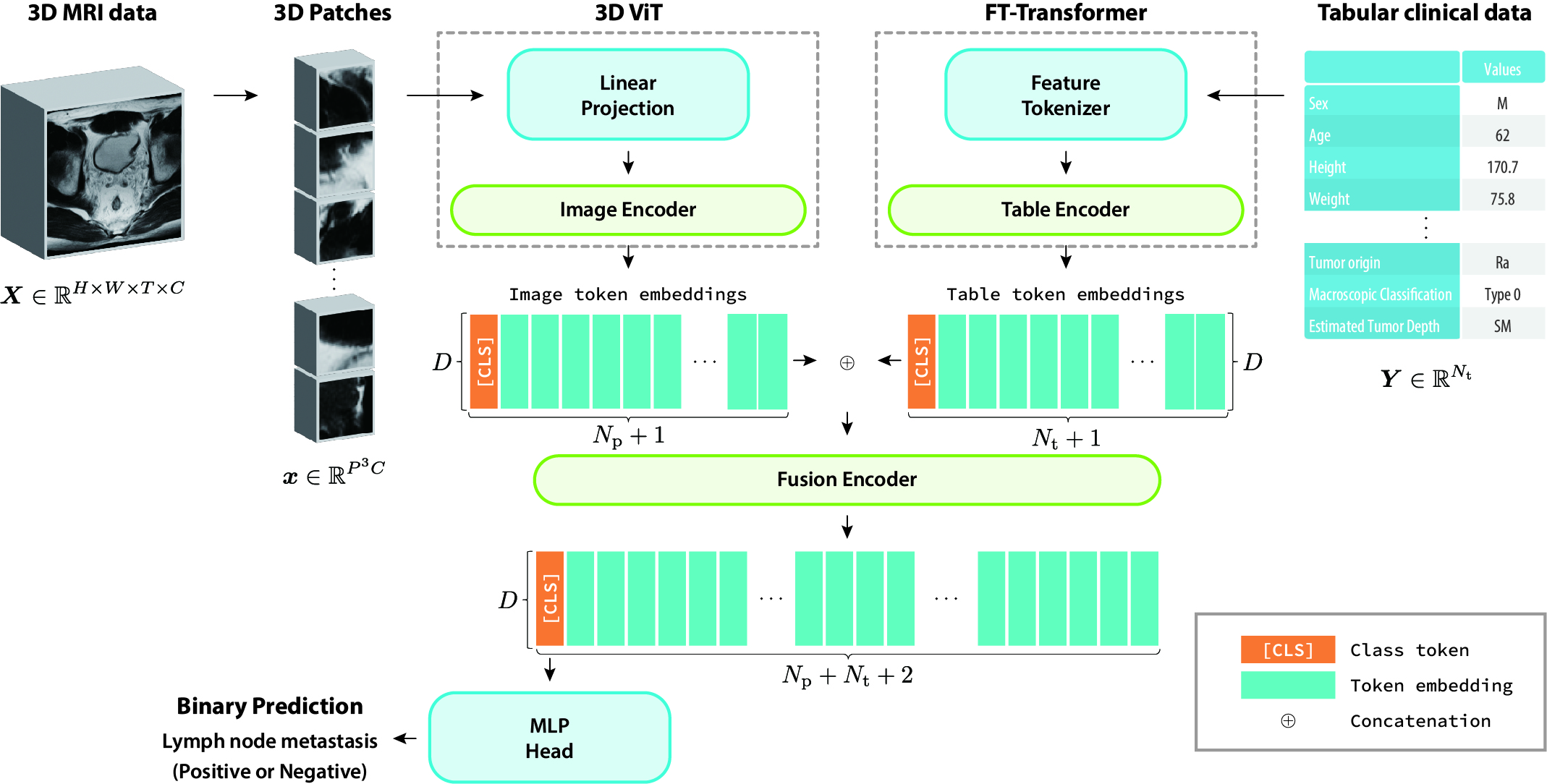}
  \caption{\textbf{The multimodal transformer architecture.} This architecture is designed for binary classification, predicting the presence or absence of postoperative pelvic lymph node metastasis using preoperative 3D MRI data and tabular clinical data. The MRI data are divided into 3D patches and then transformed into patch embeddings by the linear projection module of the 3D Vision Transformer (ViT). These patch embeddings are processed by an image encoder, resulting in \emph{image token embeddings}. In parallel, the tabular clinical data is converted into embeddings by a feature tokenizer and then processed by a table encoder within the Feature Tokenizer (FT-) Transformer, producing \emph{table token embeddings}. These two sets of embeddings are concatenated to form a combined feature vector, which is then fed into a Fusion Encoder. The leading element of the final token embeddings is input to a Multilayer Perceptron (MLP) Head. This MLP Head calculates the probability for the binary classification, determining the presence or absence of lymph node metastasis.}
  \label{fig:architecture}
\end{figure*}

\begin{table*}[t!]
\centering
\caption{\textbf{Preoperative clinical information.} LPLN, lateral pelvic lymph node; MLN, mesorectal lymph node.}
\label{tab:preoperative_clinical_data}
\resizebox{\textwidth}{!}{%
\begin{tabular}{c l l p{14cm}}
\hline
\rule{0pt}{2.6ex} % Increase the height of the row slightly
\textbf{No.} & \textbf{Attributes} & \textbf{Data Type} & \textbf{Description} \\ 
\hline
1 & Sex & Binary & Sex assigned at birth (0: female, 1: male) \\
2 & Age & Numeric & Age expressed in years \\
3 & Height & Numeric & Height measured in centimeters \\
4 & Weight & Numeric & Weight measured in kilograms \\
5 & Body Mass Index & Numeric & Body mass index calculated using height and weight \\
6 & Tumor Origin & Categorical & Tumor origin (0: rectosigmoid, 1: rectum above the peritoneal reflection, 2: rectum below the peritoneal reflection, 3: unspecified) \\
7 & Macroscopic Classification & Categorical & Macroscopic classification (0: superficial type, 1: polypoid type, 2: ulcerated type with clear margin, 3: ulcerated type with infiltration, 4: diffusely infiltrating type, 5: unclassified type) \\
8 & Estimated Tumor Depth & Categorical & Clinical assessment of tumor depth (0: mucosa, 1: submucosa, 2: muscularis propria, 3: subserosa, 4: serosa, 5: direct invasion into adjacent organs or structures, 6: unknown) \\
9 & CEA & Numerical & Carcinoembryonic antigen level measured in ng/mL \\
10 & CA19-9 & Numerical & Carbohydrate antigen level measured in U/mL \\
11 & Maximum Length of Primary Tumor & Numerical & Measured in millimeters \\
12 & Maximum Length of Right LPLN & Numerical & Measured in millimeters \\
13 & Maximum Length of Left LPLN & Numerical & Measured in millimeters \\
14 & Maximum Length of MLN & Numerical & Measured in millimeters \\
15 & Minimum Length of Primary Tumor & Numerical & Measured in millimeters \\
16 & Minimum Length of Right LPLN & Numerical & Measured in millimeters \\
17 & Minimum Length of Left LPLN & Numerical & Measured in millimeters \\
18 & Minimum Length of MLN & Numerical & Measured in millimeters \\ 
\hline
\end{tabular}
}
\end{table*}

\begin{table*}[t!]
\centering
\caption{\textbf{Detailed patient characteristics.} All numerical data are presented as mean $\pm$ standard deviation. LPLN, lateral pelvic lymph node; MLN, mesorectal lymph node.}
\label{tab:detail_dataset}
\resizebox{\textwidth}{!}{%
\begin{tabular}{llccc}
\hline
\textbf{Attributes}                                 & \textbf{Values}                                    & \multicolumn{1}{l}{\textbf{Training Dataset}} & \multicolumn{1}{l}{\textbf{Validation Dataset}} & \multicolumn{1}{l}{\textbf{Testing Dataset}} \\ \hline
Number of Patients ($N$)                                                   &                                                    & 661                                           & 60                                              & 100                                          \\ \hline
\textit{\textbf{Preoperative clinical information}} &                                                    & \multicolumn{1}{l}{}                          & \multicolumn{1}{l}{}                            & \multicolumn{1}{l}{}                         \\
Sex (assigned at birth)                             & Female                                             & 229 (34.6\%)                                  & 13 (21.7\%)                                     & 38 (38.0\%)                                  \\
                                                    & Male                                               & 432 (65.4\%)                                  & 47 (78.3\%)                                     & 62 (62.0\%)                                  \\
Age                                                 &                                                    & 60.8 $\pm$ 12.0                                   & 61.8 $\pm$ 12.1                                     & 60.9 $\pm$ 12.2                                  \\
Height                                              &                                                    & 163.1 $\pm$ 9.3                                   & 164.5 $\pm$ 7.2                                     & 162.2 $\pm$ 8.8                                  \\
Weight                                              &                                                    & 60.6 $\pm$ 12.3                                   & 61.3 $\pm$ 10.7                                     & 59.1 $\pm$ 12.5                                  \\
Body Mass Index                                     &                                                    & 22.6 $\pm$ 3.5                                    & 22.6 $\pm$ 3.6                                      & 22.3 $\pm$ 3.8                                   \\
Tumor Origin                                        & Rectosigmoid                                       & 31 (4.7\%)                                    & 4 (6.7\%)                                       & 5 (5.0\%)                                    \\
                                                    & Rectum above the peritoneal reflection             & 159 (24.0\%)                                  & 16 (26.7\%)                                     & 25 (25.0\%)                                  \\
                                                    & Rectum below the peritoneal reflection             & 462 (69.9\%)                                  & 39 (65.0\%)                                     & 70 (70.0\%)                                  \\
                                                    & Unspecified                                        & 9 (1.4\%)                                     & 1 (1.7\%)                                       & 0 (0.0\%)                                    \\
Macroscopic Classification                          & Superficial type                                   & 37 (5.6\%)                                    & 4 (6.7\%)                                       & 7 (7.0\%)                                    \\
                                                    & Polypoid type                                      & 27 (4.1\%)                                    & 1 (1.7\%)                                       & 9 (9.0\%)                                    \\
                                                    & Ulcerated type with clear margin                   & 560 (84.7\%)                                  & 53 (88.3\%)                                     & 79 (79.0\%)                                  \\
                                                    & Ulcerated type with infiltration                   & 12 (1.8\%)                                    & 1 (1.7\%)                                       & 2 (2.0\%)                                    \\
                                                    & Diffusely infiltrating type                        & 0 (0.0\%)                                     & 0 (0.0\%)                                       & 0 (0.0\%)                                    \\
                                                    & Unclassified type                                  & 25 (3.8\%)                                    & 1 (1.7\%)                                       & 3 (3.0\%)                                    \\
Estimated Tumor Depth                               & Mucosa                                             & 2 (0.3\%)                                     & 0 (0.0\%)                                       & 0 (0.0\%)                                    \\
                                                    & Submucosa                                          & 49 (7.4\%)                                    & 3 (5.0\%)                                       & 6 (6.0\%)                                    \\
                                                    & Muscularis propria                                 & 92 (13.9\%)                                   & 7 (11.7\%)                                      & 8 (8.0\%)                                    \\
                                                    & Subserosa                                          & 317 (48.0\%)                                  & 28 (46.7\%)                                     & 56 (56.0\%)                                  \\
                                                    & Serosa                                             & 130 (19.7\%)                                  & 13 (21.7\%)                                     & 22 (22.0\%)                                  \\
                                                    & Direct invasion into adjacent organs or structures & 42 (6.4\%)                                    & 2 (3.3\%)                                       & 4 (4.0\%)                                    \\
                                                    & Unknown                                            & 29 (4.4\%)                                    & 7 (11.7\%)                                      & 4 (4.0\%)                                    \\
CEA                                                 &                                                    & 15.5 $\pm$ 49.9                                   & 18.0 $\pm$ 72.5                                     & 10.1 $\pm$ 19.9                                  \\
CA19-9                                              &                                                    & 105.1 $\pm$ 1240.6                                & 24.1 $\pm$ 32.6                                     & 23.6 $\pm$ 35.7                                  \\
Maximum Length of Primary Tumor                     &                                                    & 64.1 $\pm$ 14.5                                   & 65.6 $\pm$ 16.8                                     & 63.9 $\pm$ 12.6                                  \\
Maximum Length of Right LPLN                        &                                                    & 5.7 $\pm$ 2.5                                     & 6.0 $\pm$ 3.0                                       & 5.2 $\pm$ 2.0                                    \\
Maximum Length of Left LPLN                         &                                                    & 6.1 $\pm$ 3.4                                     & 6.0 $\pm$ 2.4                                       & 5.8 $\pm$ 3.1                                    \\
Maximum Length of MLN                               &                                                    & 8.5 $\pm$ 3.8                                     & 8.8 $\pm$ 3.4                                       & 7.8 $\pm$ 3.0                                    \\
Minimum Length of Primary Tumor                     &                                                    & 56.4 $\pm$ 11.7                                   & 57.6 $\pm$ 14.1                                     & 56.1 $\pm$ 10.4                                  \\
Minimum Length of Right LPLN                        &                                                    & 4.8 $\pm$ 2.1                                     & 4.9 $\pm$ 2.5                                       & 4.5 $\pm$ 1.8                                    \\
Minimum Length of Left LPLN                         &                                                    & 5.1 $\pm$ 2.9                                     & 5.1 $\pm$ 2.3                                       & 5.0 $\pm$ 2.8                                    \\
Minimum Length of MLN                               &                                                    & 7.3 $\pm$ 3.1                                     & 7.5 $\pm$ 2.7                                       & 6.7 $\pm$ 2.5                                    \\ \hline
\textit{\textbf{Surgical procedure}}                &                                                    & \multicolumn{1}{l}{}                          & \multicolumn{1}{l}{}                            & \multicolumn{1}{l}{}                         \\
LPLN Dissection                                     & Performed                                          & 346 (52.4\%)                                  & 34 (56.7\%)                                     & 54 (54.0\%)                                  \\
                                                    & Unperformed                                        & 315 (47.6\%)                                  & 26 (43.3\%)                                     & 46 (46.0\%)                                  \\ \hline
\textit{\textbf{Pathological information}}          &                                                    & \multicolumn{1}{l}{}                          & \multicolumn{1}{l}{}                            & \multicolumn{1}{l}{}                         \\
Pathological LPLN Metastasis                        & Positive                                           & 78 (11.8\%)                                   & 7 (11.7\%)                                      & 9 (9.0\%)                                    \\
                                                    & Negative                                           & 583 (88.2\%)                                  & 53 (88.3\%)                                     & 91 (91.0\%)                                  \\
Pathological MLN Metastasis                         & Positive                                           & 314 (47.5\%)                                  & 26 (43.3\%)                                     & 41 (41.0\%)                                  \\
                                                    & Negative                                           & 347 (52.5\%)                                  & 34 (56.7\%)                                     & 59 (59.0\%)                                  \\
Pathological Pelvic Lymph Node Metastasis           & Positive                                           & 330 (49.9\%)                                  & 28 (46.7\%)                                     & 43 (43.0\%)                                  \\
                                                    & Negative                                           & 331 (50.0\%)                                  & 32 (53.3\%)                                     & 57 (57.0\%)                                  \\ \hline
\end{tabular}
}
\end{table*}

\subsection{Preprocessing of MRI data}\label{sec:preprocess}

From the collected MRI data, we selected T2-weighted images due to their superior contrast in the tumor region. In the preprocessing, the voxel intensities of each T2-weighted image were normalized using Z-score normalization, and voxel sizes were resampled to a normalized size of (0.72, 0.72, 3.0) for (height, width, thickness). Note that each T2-weighted image has a different volume extension. We then cropped a three-dimensional (3D) volume from each image, centered around the bounding box encompassing each primary tumor site. The size of the cropped volume is denoted as $\bm{X} \in \mathbb{R}^{H \times W \times T \times C}$, where $H$, $W$, and $T$ represent the axial height, axial width, and longitudinal thickness (i.e., depth) of the volume, respectively, and $C$ is the number of channels.

\subsection{Designing Multimodal Transformer}\label{sec:multimodal_transformer}

A multimodal transformer, designed to predict pathological lymph node metastasis from tabular clinical data and MRI scans, was developed (see \textbf{\cref{fig:architecture}}). To accommodate its multimodal input, we combined three transformer-based architectures in its design: a \emph{3D ViT} for the MRI data, a \emph{Feature Tokenizer (FT-) Transformer} for the tabular clinical data, and a \emph{Fusion Encoder} to integrate the information from both sources for the final prediction.

\subsubsection{3D Vision Transformer for MRI data}\label{sec:vit_encoder}

A 3D ViT-based architecture \citep{Alexey2021, Lahoud20223}, which comprises a linear projection module and an image encoder module, is used for encoding imaging information from T2-weighted images.

First, the linear projection module divides the input volume $\bm{X} \in \mathbb{R}^{H \times W \times T \times C}$ into $(H / P, W / P, T / P)$ patches, denoted as $\bm{P} \in \mathbb{R}^{P \times P \times P \times C}$. Subsequently, it flattens them into one-dimensional vectors, $\bm{x} \in \mathbb{R}^{P^3C}$. As a result, a sequence of these flattened patches $[\bm{x}_1, ..., \bm{x}_{N_\mathrm{p}}]$ is generated, where $N_\mathrm{p} = HWT / P^3$ represents the number of patches. These flattened patches are then mapped to a $D$-dimensional latent space using a trainable linear projection to produce patch embeddings $[\bm{z}_1, ..., \bm{z}_{N_\mathrm{p}}] \in \mathbb{R}^{N_\mathrm{P} \times D}$. A class token of the same dimension, $\bm{z}_\mathrm{c} \in \mathbb{R}^{D}$, is concatenated to the beginning of the sequence, and learnable position embeddings $\bm{E}_\mathrm{pos} \in \mathbb{R}^{(N_\mathrm{p} + 1) \times D}$ are added to the patch embeddings to retain the positional information of the patches \citep{Alexey2021}. The initial sequence of token embeddings is thus given by $\bm{Z}_0 = [\bm{z}_\mathrm{c}, \bm{z}_1, ..., \bm{z}_{N_\mathrm{p}}] + \bm{E}_\mathrm{pos} \in \mathbb{R}^{(N_\mathrm{p} + 1) \times D}$.

The initial sequence of token embeddings is then processed by the image encoder. The image encoder consists of $L$ encoder blocks, each comprising a multi-head self-attention (MHSA) module and a multilayer perceptron (MLP) module. The MHSA module has $H$ heads, and the MLP module has a $F$-dimensional intermediate layer. The output of the $l$-th encoder block $\bm{Z}_l \in \mathbb{R}^{(N_\mathrm{p} + 1) \times D}$ can be formulated as follows:
\begin{equation}\label{eq:encoder_block}
\bm{z}^{\prime}_{l} = \mathrm{MHSA}(\mathrm{LN}(\bm{z}_{l-1})) + \bm{z}_{l-1},
\end{equation}
\begin{equation}\label{eq:encoder_block_2}
\bm{z}_{l} = \mathrm{MLP}(\mathrm{LN}(\bm{z}^{\prime}_{l})) + \bm{z}^{\prime}_{l},
\end{equation}
where LN represents layer normalization. Finally, $\bm{Z}_L \in \mathbb{R}^{(N_\mathrm{p} + 1) \times D}$ is referred to as the \emph{image token embeddings}.

\subsubsection{FT-Transformer for Tabular Clinical Data}\label{sec:ft_transformer_encoder}

A network architecture based on FT-Transformer \citep{Gorishniy2023}, which consists of a feature tokenizer module and a table encoder module, is used to process tabular clinical data. The FT-Transformer was originally introduced as a straightforward adaptation for managing tabular data, effectively transforming both categorical and numerical features into embeddings.

The feature tokenizer module initially converts the input feature with $N_\mathrm{t}$ variables, denoted as $\bm{Y} \in \mathbb{R}^{N_\mathrm{t}}$, into an embedding $[\bm{w}_1, ..., \bm{w}_{N_\mathrm{t}}] \in \mathbb{R}^{N_\mathrm{t} \times D}$. During this process, element-wise multiplication with the vector $\bm{M}^{\mathrm{(num)}} \in \mathbb{R}^{D}$ is applied to the $j$-th numerical variable $y_j$. Conversely, for the $j^{\prime}$-th categorical variable $y_{j^{\prime}}$, a lookup table $\bm{M}^{\mathrm{(cat)}} \in \mathbb{R}^{n_{j^{\prime}} \times D}$ is used to convert it into an embedding $\bm{w}_{j^{\prime}}$, where $n_{j^{\prime}}$ represents the number of categories of the corresponding feature. Subsequently, a class token $\bm{w}_\mathrm{c} \in \mathbb{R}^{D}$ is concatenated to the beginning of the sequence, forming the initial sequence of the token embeddings as: $\bm{W}_0 = [\bm{w}_\mathrm{c}, \bm{w}_1, ..., \bm{w}_{N_\mathrm{t}}] \in \mathbb{R}^{(N_\mathrm{t} + 1) \times D}$. The table encoder then processes this initial sequence, outputting the \emph{table token embeddings}, $\bm{W}_L \in \mathbb{R}^{(N_\mathrm{t} + 1) \times D}$, through $L$ encoder blocks. These blocks consist of MHSA and MLP modules, similar to those in the image encoder in 3D ViT.

\subsubsection{Fusion Encoder for the Multimodal Integration}\label{sec:fusion_encoder}

The Fusion Encoder is a transformer-based architecture, akin to both the image and table encoders. This encoder processes a concatenated feature vector $\bm{V}_0 \in \mathbb{R}^{(N_\mathrm{p} + N_\mathrm{t} + 2) \times D}$, which combines image token embeddings, $\bm{Z}_L$, with table token embeddings, $\bm{W}_L$. It processes these vectors through $L$ encoder blocks equipped with MHSA and MLP modules. The first element of the resultant feature vector $\bm{V}_L$ is then input to an MLP head, which calculates the probability of pathological pelvic lymph node metastasis.

\subsection{Training of Multimodal Transformer}\label{sec:training_multimodal_transformer}

The training of the multimodal transformer was carried out in two phases: the \emph{pre-training phase} and the \emph{training phase}.

\subsubsection{Learning Objectives of Pre-training Phase}\label{sec:learning_objectives_pretrain}

During the initial pre-training phase (see \textbf{step 1} in \textbf{\cref{fig:dataset}}), the 3D ViT was pre-trained using the pre-training dataset that included annotations of bounding boxes around primary tumors. Specifically, a volume of the same dimensions as the input volume was extracted from each T2-weighted image, centered within the bounding box. This extracted volume was designated as the \emph{anchor volume}, represented as $\bm{X} \in \mathbb{R}^{H \times W \times T \times C}$. Subsequently, another volume was cropped randomly from within the bounding box, termed the \emph{positive volume}, $\bm{X}^{+} \in \mathbb{R}^{H \times W \times T \times C}$, and a volume cropped randomly from an area outside the bounding box was labeled as the \emph{negative volume}, $\bm{X}^{-} \in \mathbb{R}^{H \times W \times T \times C}$. In essence, the anchor and positive volumes contain imaging information proximal to the primary tumor, while the negative volume encompasses anatomical information from more distant areas. Finally, the pre-training aimed to minimize the triplet margin loss \citep{Balntas2016}, using image token embeddings $\bm{Z}_L$ derived from the triplet $(\bm{X}, \bm{X}^{+}, \bm{X}^{-})$. This is mathematically represented as:
\begin{equation}\label{eq:triplet}
L_\mathrm{margin} = \max \{d(f(\bm{X}), f(\bm{X}^+)) - d(f(\bm{X}), f(\bm{X}^-)) + 1, 0\},
\end{equation}
where $d(\cdot , \cdot)$ represents the L2 distance between two image token embeddings, and $f$ signifies the function of the 3D ViT.

\subsubsection{Learning Objectives of Training Phase}\label{sec:learning_objectives}

We trained the pre-trained 3D ViT in conjunction with the FT-Transformer and Fusion Encoder for binary classification of pelvic lymph node metastasis, using the training dataset (see \textbf{step 2} in \textbf{\cref{fig:dataset}}). We monitored its performance using the validation dataset. Both datasets include MRI data, tabular clinical data, and postoperative pathological information. The MRI data is input into the 3D ViT, while the tabular clinical data is input into the FT-Transformer. On the other hand, postoperative pathological information provided details on the presence or absence of metastasis in both the left and right LPLNs and MLNs. Cases with metastasis in any of the left or right LPLNs or MLNs were classified as ``positive'' for pelvic lymph node metastasis, while those without metastasis in these areas were classified as ``negative''. The loss function was defined as a weighted combination of binary cross-entropy (BCE) $L_\mathrm{BCE}$ and cosine loss $L_\mathrm{cos}$, expressed as:
\begin{equation}\label{eq:training_objective}
L = w_\mathrm{BCE} L_\mathrm{BCE} + w_\mathrm{cos} L_\mathrm{cos}.
\end{equation}

Cosine loss is known to enhance performance when combined with cross-entropy loss, particularly in small datasets \citep{Barz2020}. Through the hyperparameter search, as described in \textbf{\cref{sec:implementation_details}}, we observed its beneficial effect on model performance in our case study.

\subsection{Inference with Uncertainty Estimation}\label{sec:inference_uncertainty}

In the inference stage (see \textbf{step 3} in \textbf{\cref{fig:dataset}}), test-time data augmentation was utilized to calculate the probability of the positive class regarding pelvic lymph node metastasis and its variance, which can serve as the basis for estimating model uncertainty \citep{Wang2019}. A volume $\bm{X} \in \mathbb{R}^{H \times W \times T \times C}$ is cropped around a randomly chosen center point, within a range of $K$ pixels from the center of the bounding box, and a horizontal flip is applied with a probability of 0.5. This transformed volume $\bm{X}^{\prime} \in \mathbb{R}^{H \times W \times T \times C}$ is then fed into the multimodal transformer to calculate the positive class probability. This process is repeated $Q$ times for each sample, after which the mean, maximum, and variance of the positive class probabilities are computed.

Based on the prediction performance comparison in the validation dataset, the maximum of the positive class probabilities, rather than the mean, was used as the inferred class probability $\omega_\mathrm{prob}$. When the class probability exceeds 0.5 ($\omega_\mathrm{prob} \geq 0.5$), lymph node metastasis is predicted to be positive; conversely, if it is below 0.5 ($\omega_\mathrm{prob} < 0.5$), it is considered negative. Moreover, the variance of the positive class probability, denoted as $\sigma_\mathrm{prob}$, was employed to determine the uncertainty score as a piece of meta-level information (see \textbf{\cref{sec:formulation_uncertainty_estimator}}).

\subsection{Visualizing Attention Maps}\label{sec:attention_visualization}

When visualizing the attention map at the inference stage, the input volume $\bm{X} \in \mathbb{R}^{H \times W \times T \times C}$ is cropped so that the center of the bounding box aligns with the center of the input volume. This visualization method is based on the implementation of Transformer Explainability as described by \cite{Chefer2021}. The attention map is computed separately for the MRI data and the tabular clinical data, with attention values normalized such that the maximum value is 1. These visualized attention maps are then evaluated by expert physicians to compute the agreement score (refer to \textbf{\cref{sec:agreement_score_mri}} and \textbf{\cref{sec:agreement_score_tabular}}).

\subsection{Implementation Details}\label{sec:implementation_details}

In implementing the multimodal transformer, we determined the hyperparameters as follows: the input volume are set to $H = 192$, $W = 192$, $T = 64$, and $C=1$; for the 3D ViT, the settings are $P = 16$, $D = 1024$, $H = 16$, $F = 2048$, and $L = 6$; for the FT-Transformer, they are $D = 1024$, $H = 8$, $F = 4096$, and $L = 6$; and for the Fusion Encoder, they are $D = 1024$, $H = 8$, $F = 2048$, and $L = 6$. The intermediate layer of the MLP head has a dimension of $512$.

For pre-training, we use a batch size of $80$, a number of epochs $= 300$, a learning rate $= 3.0 \times 10^{-5}$ with the Adam optimizer \citep{Kingma2017}, a weight decay $= 5.0 \times 10^{-4}$, and data augmentation functions [RandomBiasField, RandomHorizontalFlip]. For training, we maintain the same batch size, number of epochs, and learning rate, along with $w_\mathrm{BCE} = 1.0$, $w_\mathrm{cos} = 10.0$, and data augmentation functions [RandomBiasField, RandomCropping, RandomHorizontalFlip]. The RandomCropping function crops the input volume, centering it within the bounding box of the primary tumor. The dropout probability, denoted as $p_\mathrm{drop}$, is set at 0.1 in the 3D ViT, FT-Transformer, and Fusion Encoder, and at 0.2 in the MLP head. Label smoothing \citep{Müller2020} was employed to mitigate overfitting to the training dataset in the multimodal transformer. For the inference with uncertainty estimation, $K = 100$ and $Q = 100$ were set for the test-time data augmentation. 

These hyperparameters were established through a grid search, considering the following candidate values: $P = \{8, 16, 32\}$, learning rate in the training phase $= \{1.0 \times 10^{-5}, 3.0 \times 10^{-5}, 5.0 \times 10^{-5}, 8.0 \times 10^{-5}, 1.0 \times 10^{-4}, 5.0 \times 10^{-4}, 1.0 \times 10^{-3}\}$, weight decay in the training phase $= \{1.0 \times 10^{-5}, 1.0 \times 10^{-4}, 1.0 \times 10^{-3}\}$, $w_\mathrm{BCE} = \{0, 1, 5, 10, 20\}$, $w_\mathrm{cos} = \{0, 1, 5, 10, 20\}$, and $p_\mathrm{drop} = \{0.1, 0.2, 0.5, 0.7\}$. 

\section{Evaluation Methods}\label{sec:evaluation}

After the training and inference phases with the multimodal transformer, we established uncertainty estimators to demonstrate the beneficial effect of model explainability on model reliability (see \textbf{\cref{fig:question}b}). It is also important to note that the \emph{correctness} of each prediction was calculated, with a value of 0 assigned for incorrect predictions and 1 for correct ones, to facilitate subsequent analysis. Hereafter, in discussions involving two machine-learning models -- the multimodal transformer and the uncertainty estimator -- the term ``model'' will specifically refer to the former.

\subsection{Prediction Performance}\label{sec:prediction_performance}

For the evaluation of the prediction capacity, the \emph{accuracy}, \emph{precision}, \emph{recall}, \emph{F1-score}, and \emph{Area Under the Receiver Operating Characteristic Curve} (AUC-ROC) were calculated based on the ground truth obtained from postoperative pathological information and the prediction results from the multimodal transformer. These metrics were calculated in both the validation and testing datasets (see \textbf{step 3} in \textbf{\cref{fig:dataset}}).

\subsection{Agreement Score for MRI Data}\label{sec:agreement_score_mri}

\begin{figure}[t!]
  \centering
  \includegraphics[width=\columnwidth]{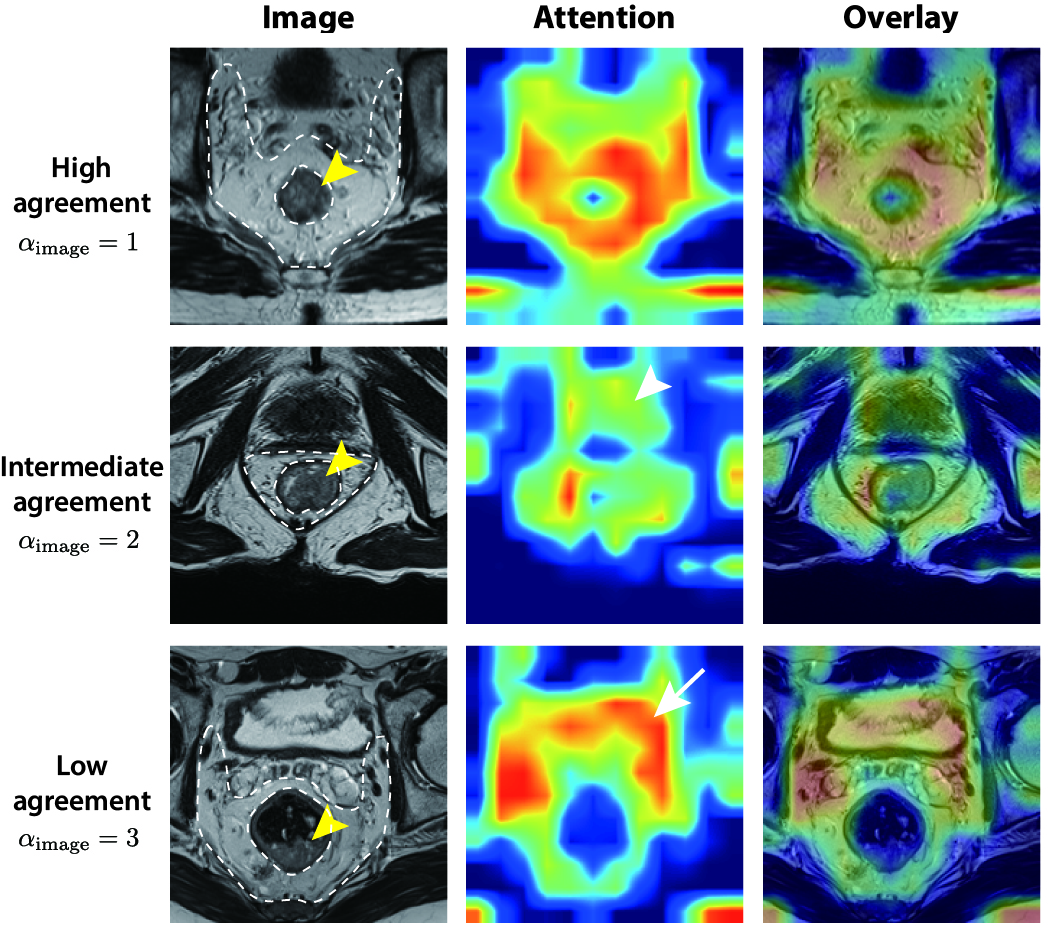}
  \caption{\textbf{Agreement scores for MRI data.} Consensus-based scoring by experts was conducted for MRI data with visualized attention maps. The pelvic lymph node area corresponds to the inside enclosed by two white dotted lines. \textbf{(Top)} High agreement was judged for a case where attention was accumulated preferentially in the pelvic lymph node area. \textbf{(Middle)} Intermediate agreement was noted for a case in which attention was equally distributed not only in the pelvic lymph node area but also in another anatomical region (see white arrowhead). \textbf{(Bottom)} Low agreement was determined for a case where attention to the other area (see white arrow) was more prominent than to the pelvic lymph node area. In all cases, the yellow arrowheads indicate the primary lesions.}
  \label{fig:image_attention}
\end{figure}

Assessment of the agreement score for the visualized attention maps on MRI data was conducted (see \textbf{step 4} in \textbf{\cref{fig:dataset}}). The scoring was carried out for all samples in both the validation dataset ($N = 60$) and the testing dataset ($N = 100$), without informing the evaluators of the actual pathological information for each case.

For assessing the agreement score, three experts with over 10 years of experience (a radiologist, a colorectal surgeon, and a radiation oncologist) participated as evaluators. As shown in \textbf{\cref{fig:image_attention}}, the evaluators conducted a consensus-based scoring on the attention maps visualized in the images, assigning scores of 1, 2, or 3 to each image. The criteria for the scoring are as follows:
\begin{enumerate}
\item \textbf{High agreement:} Preferential accumulation of attention in the pelvic lymph node area.
\item \textbf{Intermediate agreement:} Accumulation of attention in the pelvic lymph node area, but with a similar level of attention in other anatomical regions.
\item \textbf{Low agreement:} More prominent accumulation of attention in anatomical regions other than the pelvic lymph nodes.
\end{enumerate}
The pelvic lymph node area is the anatomical location where physicians generally focus their attention to diagnose pelvic lymph node enlargement before rectal cancer surgery. Therefore, our scoring reflects the consistency between physician understanding and model explainability, defined as the agreement score for the MRI data, denoted as $\alpha_\mathrm{image}$. The Pearson correlation coefficient was used to evaluate the correlation between the model's prediction correctness and the agreement score.

\subsection{Agreement Score for Tabular Clinical Data}\label{sec:agreement_score_tabular}

For assessing the agreement score for tabular clinical data (see \textbf{step 4} in \textbf{\cref{fig:dataset}}), we employed a ranking-based comparison between the \textbf{physician-assessed ranking} and the \textbf{attention-based ranking} (see \textbf{\cref{fig:table_attention}}). To prepare the physician-assessed ranking, the evaluators ranked the 18 items of preoperative clinical information according to their relevance to postoperative pelvic node metastasis, based on consensus. This ranking is representative of the physician’s understanding. Conversely, for the attention maps in both the validation and testing datasets, the ranking of the 18 preoperative clinical information items is computed based on the magnitude of attention each item receives. Spearman's rank correlation coefficient, a non-parametric measure of rank correlation, is calculated based on the physician-assessed ranking and the attention-based ranking for each sample. The resulting correlation coefficient is considered as the agreement score for the tabular clinical data, denoted by $\alpha_\mathrm{table}$. Additionally, the Pearson correlation coefficient was utilized to assess its correlation with the correctness of the model's predictions.

\subsection{Formulation of Uncertainty Estimators}\label{sec:formulation_uncertainty_estimator}

In our framework (see \textbf{\cref{fig:question}a}), a good explanation is deemed capable of providing additional meta-level information that reduces overall uncertainty through agreement with physician understanding. To quantify this, we utilized a machine learning-based uncertainty estimator that estimates the normalized form of model uncertainty, referred to as the uncertainty score (see \textbf{\cref{fig:question}b}), which is calculated based on the correctness of each prediction. Note that the uncertainty estimator is categorized into two types: \textbf{uninformed estimator} and \textbf{informed estimator}, depending on the use of additional information derived from the agreement between model explainability and physician understanding. The difference in performance between these two estimators can be interpreted as indicative of the impact that such information has on model reliability.

\subsubsection{Uninformed Estimator}\label{sec:uninformed}

The uninformed estimator $\mathcal{U}(\sigma_\mathrm{prob})$ predicts the uncertainty score based solely on the variance of the positive class probability, $\sigma_\mathrm{prob}$ (see \textbf{\cref{sec:inference_uncertainty}}). This estimation corresponds to the model's uncertainty before incorporating any additional information derived from the agreement with physician understanding. The formulation is as follows:

\begin{equation}\label{eq:uninformed_model}
\begin{alignedat}{4}
\mathcal{U}(\sigma_\mathrm{prob}) &= 1 - p_\mathrm{correct}, \\
&= u_\mathrm{uninformed},
\end{alignedat}
\end{equation}
where $p_\mathrm{correct}$ represents the probability of correctness of the model's prediction, and $u_\mathrm{uninformed}$ is the uncertainty score calculated by the uninformed estimator.

\subsubsection{Informed Estimator}\label{sec:informed}

The informed estimator $\mathcal{U}(\sigma_\mathrm{prob}, \alpha_\mathrm{image}, \alpha_\mathrm{table})$ predicts the uncertainty score by incorporating additional information, including the agreement scores for MRI data $\alpha_\mathrm{image}$ (see \textbf{\cref{sec:agreement_score_mri}}) and for tabular clinical data $\alpha_\mathrm{table}$ (see \textbf{\cref{sec:agreement_score_tabular}}). The estimation can be interpreted as reflecting the model's uncertainty, enriched with agreement scores that provide insights into the model's explainability. This is mathematically formulated as follows:

\begin{equation}\label{eq:informed_model}
\begin{alignedat}{4}
\mathcal{U}(\sigma_\mathrm{prob}, \alpha_\mathrm{image}, \alpha_\mathrm{table}) &= 1 - p_\mathrm{correct}, \\
&= u_\mathrm{informed},
\end{alignedat}
\end{equation}
where $p_\mathrm{correct}$ represents the probability of correctness of the model's prediction, and $u_\mathrm{informed}$ is the uncertainty score calculated by the informed estimator.

\subsection{Training of Uncertainty Estimators}\label{sec:training_uncertainty_estimator}

For the modeling function, we employed logistic regression. In the preprocessing step, Z-score normalization was performed on the variance of the positive class probability $\sigma_\mathrm{prob}$ and the agreement scores for tabular clinical data $\alpha_\mathrm{table}$, based on the statistical properties in the validation dataset. For the agreement scores for MRI data $\alpha_\mathrm{image}$, min-max normalization was applied. Both uninformed and informed estimators were trained using the validation dataset (see \textbf{step 5} in \textbf{\cref{fig:dataset}}).

\subsection{Evaluation based on the Inference of Uncertainty Estimators}\label{sec:inference_uncertainty_estimator}

After establishing the uncertainty estimators, we quantitatively assessed the utility of additional explanations in reducing overall uncertainty. This assessment involved comparing the uncertainty scores of both uninformed and informed estimators within the testing dataset (see \textbf{step 6} in \textbf{\cref{fig:dataset}}). Specifically, we examined the Pearson correlation coefficient to assess the relationship between the uncertainty scores and the actual correctness of predictions. Additionally, we analyzed how these uncertainty scores corresponded with the prediction performance of the multimodal transformer. This integrated analysis finally enabled us to observe variations in the model's predictive performance at different levels of tolerable uncertainty.

\begin{figure}[t!]
  \centering
  \includegraphics[width=\columnwidth]{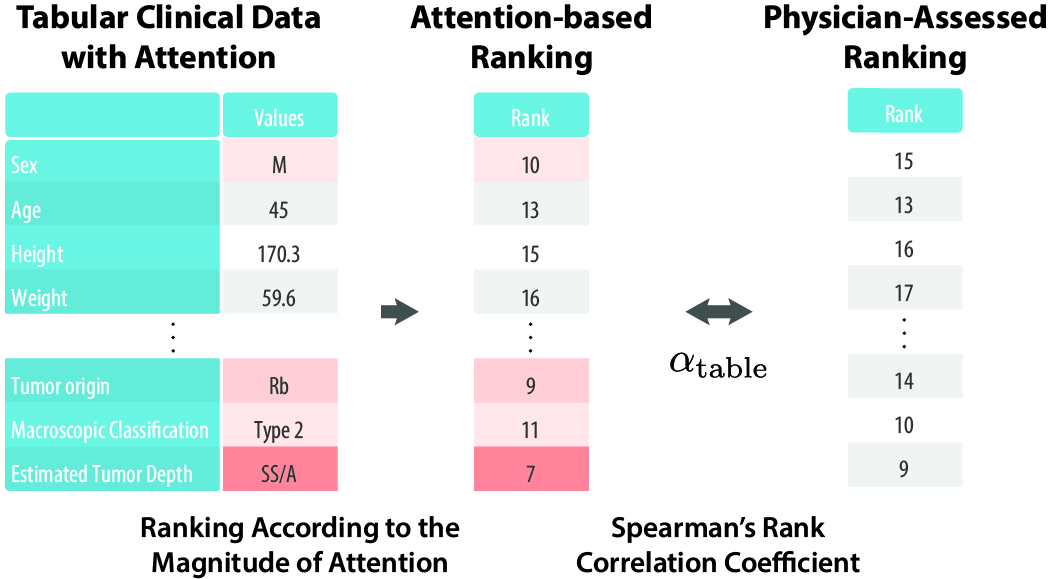}
  \caption{\textbf{Agreement scores for tabular clinical data.} For tabular clinical data with attention, an attention-based ranking was generated based on the magnitude of the attention values that items received. Conversely, the physician-assessed ranking was prepared based on the consensus among expert physicians. Subsequently, Spearman's rank correlation coefficient was computed between these rankings to calculate the agreement score for the tabular clinical data.}
  \label{fig:table_attention}
\end{figure}

\section{Results}\label{sec:results}

After demonstrating the overall prediction performance of the multimodal transformer, we document the results regarding the agreement between the visualized attention maps and physician understanding. Thereafter, we describe how this agreement affects the uncertainty estimator, which is essential for revealing the practical benefits of model explainability within our framework (see \textbf{\cref{fig:question}}).

\subsection{Overall Predictive Performance of the Multimodal Transformer}\label{sec:result_predictive_performance}

The overall predictive performance of the multimodal transformer for pelvic lymph node metastasis was as follows (see \textbf{step 3} in \textbf{\cref{fig:dataset}}): In the validation dataset ($N = 60$), the accuracy was $0.70$, precision was $0.63$, recall was $0.86$, F1-score was $0.73$, and AUC-ROC value was $0.78$. In the testing dataset ($N = 100$), the accuracy was $0.60$, precision was $0.53$, recall was $0.72$, F1-score was $0.61$, and AUC-ROC value was $0.68$. 

\subsection{Evaluation of Visualized Attention Maps on the MRI Data}\label{sec:result_agreement_mri_data}

As described in \textbf{\cref{sec:agreement_score_mri}}, the evaluators consisting of three experts scored the visualized attention maps in each MRI to assess whether they were distributed in a clinically reasonable manner (see \textbf{step 4} in \textbf{\cref{fig:dataset}}). This scoring is defined as the agreement score for the MRI data, $\alpha_\mathrm{image}$, which ranges from 1 to 3, with 1 being the best and 3 being the worst. The resultant frequency of the scores were as follows: In the validation dataset, the proportions were 1: $16.7\%$, 2: $50.0\%$, 3: $33.3\%$; in the testing dataset, they were 1: $18.0\%$, 2: $44.0\%$, 3: $38.0\%$. \textbf{\cref{fig:image_attention}} displays examples of scored MRI data evaluated in the testing dataset.

When the attention map in MRI data aligns with physician understanding, it can be interpreted that the multimodal transformer is focusing on clinically meaningful areas for predicting lymph node metastasis in the pelvis. Thus, it is beneficial to investigate whether there is a correlation between the agreement score and the actual correctness (whether each prediction is accurate or not). Based on this consideration, we calculated the Pearson correlation coefficient between these agreement scores and the correctness of each prediction. The coefficient was $-0.05$ ($p = 0.69$) in the validation dataset and $0.03$ ($p = 0.78$) in the testing dataset. Neither of these coefficients was statistically significant, as indicated by their $p$-values being greater than $0.05$. Consequently, the agreement score for MRI data alone did not show a significant correlation with the correctness of the model's predictions.

\subsection{Evaluation of Attention Maps in Tabular Clinical Data}\label{sec:result_agreement_clinical_data}

Subsequently, we determined the correlation between the attention-based ranking, which was derived from the magnitude of attention calculated for the tabular clinical data of each sample, and the physician-assessed ranking, using Spearman's rank correlation coefficient (see \textbf{\cref{sec:agreement_score_tabular}} and \textbf{step 4} in \textbf{\cref{fig:dataset}}). This Spearman's rank correlation coefficient is defined as the agreement score for the tabular clinical data, $\alpha_\mathrm{table}$ (see \textbf{\cref{fig:table_attention}}). The mean $\pm$ standard deviations and the range of this agreement score are as follows: $0.10 \pm 0.22$ (range: $-0.31$ -- $0.46$) in the validation dataset and $0.11 \pm 0.25$ (range: $-0.49$ -- $0.65$) in the testing dataset.

We conducted an investigation to determine if a correlation exists between the agreement score for tabular clinical data and the correctness of the model's predictions. When measuring the Pearson correlation coefficient between this agreement score and the correctness of each prediction, it was $0.11$ ($p = 0.41$) in the validation dataset and $-0.03$ ($p = 0.77$) in the testing dataset, with neither showing statistically significant correlations ($p > 0.05$). Therefore, akin to the agreement score for MRI data, the agreement on the attention maps in tabular clinical data alone did not demonstrate any significant correlation with the correctness of the model's predictions.

\begin{figure*}[t!]
  \centering
  \includegraphics[width=\textwidth]{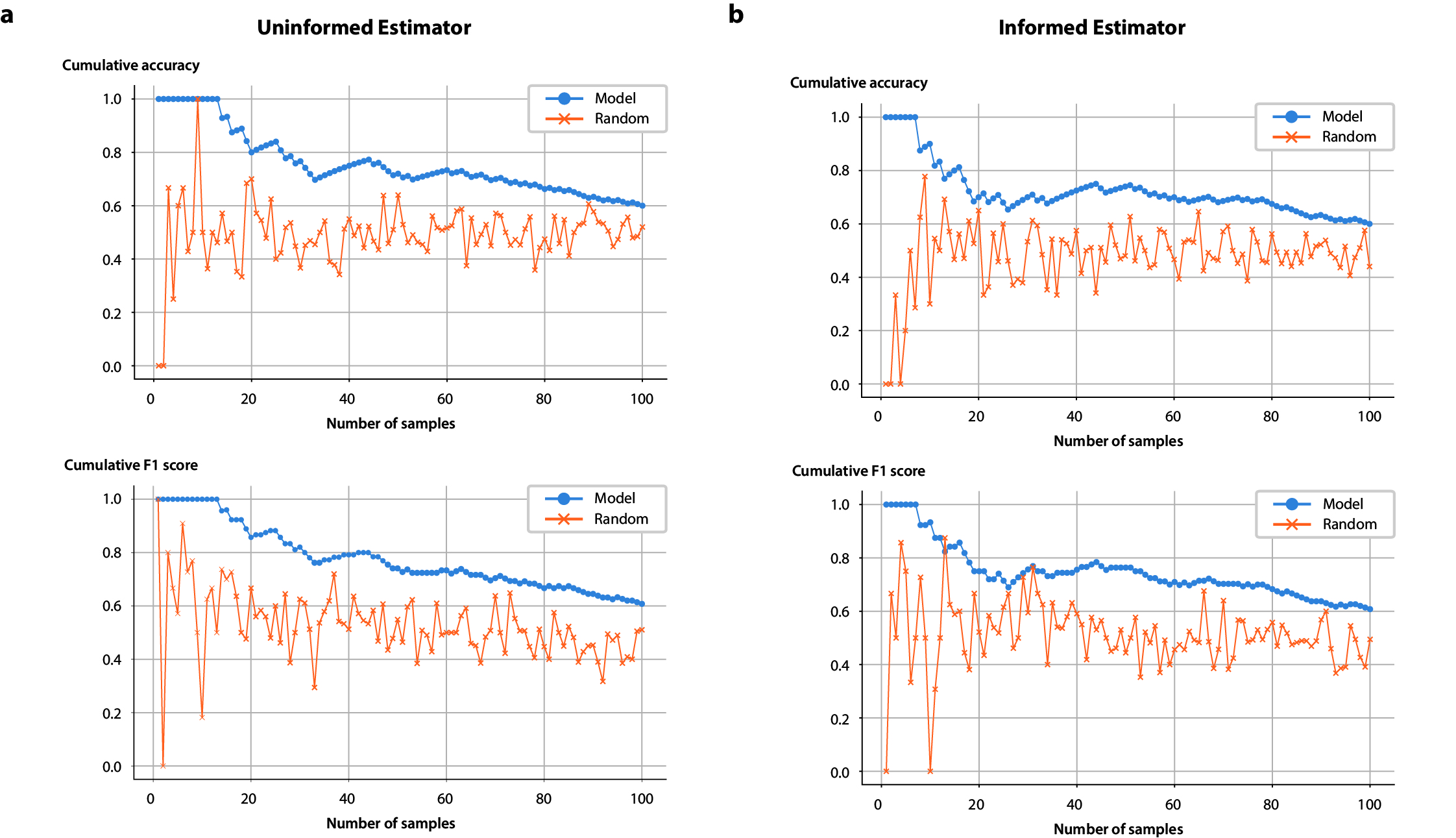}
  \caption{\textbf{Cumulative performance metrics as the function of the retaining samples based on the ascending uncertainty scores.} The vertical axis represents cumulative metrics (upper: accuracy and lower: F1-score), and the horizontal axis shows the retained number of samples, arranged in order of ascending uncertainty score. For both \textbf{(a)} the uninformed estimator $\mathcal{U}(\sigma_\mathrm{prob})$ and \textbf{(b)} the informed estimator $\mathcal{U}(\sigma_\mathrm{prob}, \alpha_\mathrm{image}, \alpha_\mathrm{table})$, the cumulative metrics exhibit a decreasing trend as the number of retained samples increases, corresponding with higher uncertainty scores. This demonstrates that a lower uncertainty score is associated with higher predictive correctness, enabling physicians to select more reliable predictions based on the uncertainty score. Notably, this relationship did not change significantly between the uncertainty scores from the uninformed and informed estimators, implying that the additional information derived from the agreement did not have a profound effect on the model's uncertainty. In each graph, ``Random'' denotes the results achieved under completely random predictions.}
  \label{fig:result_uncertainty_estimator}
\end{figure*}

\subsection{Performance of the Uninformed Model for Estimating Uncertainty Scores}\label{sec:performance_base_model}

Based on the methods described in \textbf{\cref{sec:formulation_uncertainty_estimator}} and \textbf{\cref{sec:training_uncertainty_estimator}}, we trained the uninformed estimator $\mathcal{U}(\sigma_\mathrm{prob})$ to generate the uncertainty score for each prediction, denoted as $u_\mathrm{uninformed}$ (see \textbf{step 5} in \textbf{\cref{fig:dataset}}). These uncertainty scores were derived from the variance of the positive class probability, $\sigma_\mathrm{prob}$, which serves as the meta-level information for the multimodal transformer. This approach establishes a baseline for the model's uncertainty \emph{without} incorporating additional information regarding the agreement between the model's explanation and physician understanding. It is important to note that in this context, the term ``model'' specifically refers to the multimodal transformer and not the uncertainty estimator.

In the testing dataset, we sorted the samples in ascending order of their uncertainty scores (refer to \textbf{step 6} in \textbf{\cref{fig:dataset}}). We then plotted the cumulative accuracy and F1-score variations as a function of the number of retained samples, which were ordered by ascending uncertainty scores. This approach enabled us to visualize how changes in predictive performance correlated with increases in the model's uncertainty. As depicted in \textbf{\cref{fig:result_uncertainty_estimator}a}, the cumulative accuracy and F1-score generally followed a decreasing trend. This indicates that an increase in the uncertainty score is associated with a decrease in predictive performance.

Subsequently, we calculated the Pearson correlation coefficients between the uncertainty scores and the actual correctness of the multimodal transformer in the testing dataset. The correlation coefficient was $-0.30$ ($p = 0.003$), thereby statistically confirming a mildly negative correlation ($p < 0.05$). Consequently, samples with lower uncertainty scores tended to exhibit higher predictive correctness, and vice versa. This indicates that the uncertainty score, derived from the uninformed model, serves as a useful indicator for physicians to discern which predictions are more reliable.

\subsection{Impact of the Agreement with Physician Understanding on the Model Uncertainty}\label{sec:performance_extended_model}

Furthermore, we investigated the impact of the agreement between model explanations and physician understanding on the calculation of uncertainty scores, using the informed model $\mathcal{U}(\sigma_\mathrm{prob}, \alpha_\mathrm{image}, \alpha_\mathrm{table})$ (refer to \textbf{\cref{sec:informed}}). This estimation takes into account the overall uncertainty \emph{with} additional information pertaining to the agreement. We trained informed models on the validation dataset, utilizing three pieces of meta-level information: the variance of the positive class probability $\sigma_\mathrm{prob}$, the agreement score for MRI data $\alpha_\mathrm{image}$, and the agreement score for tabular clinical data $\alpha_\mathrm{table}$.

Similar to the approach with uninformed models, we first plotted cumulative metrics, such as accuracy and F1-score, against the number of retained samples. These samples were ordered based on ascending uncertainty scores. As depicted in \textbf{\cref{fig:result_uncertainty_estimator}b}, the generally decreasing relationship observed is akin to that observed in the uninformed estimator. Notably, the Pearson correlation coefficient with actual correctness was $-0.28$ ($p = 0.006$), indicating statistical significance ($p < 0.05$). However, compared to the uninformed model, the magnitude of the correlation coefficient was slightly lower and did not outperform as expected (uninformed estimator: $-0.30$ vs. informed estimator: $-0.28$). If the agreement between model explainability and physician understanding were to contribute beneficial additional information, the correlation coefficient in the informed estimator should be less than $-0.30$. However, this was not the case. Consequently, we found no significant effect of the agreement between model explainability and physician understanding on estimating the overall model uncertainty.

\begin{figure}[t!]
  \centering
  \includegraphics[width=\columnwidth]{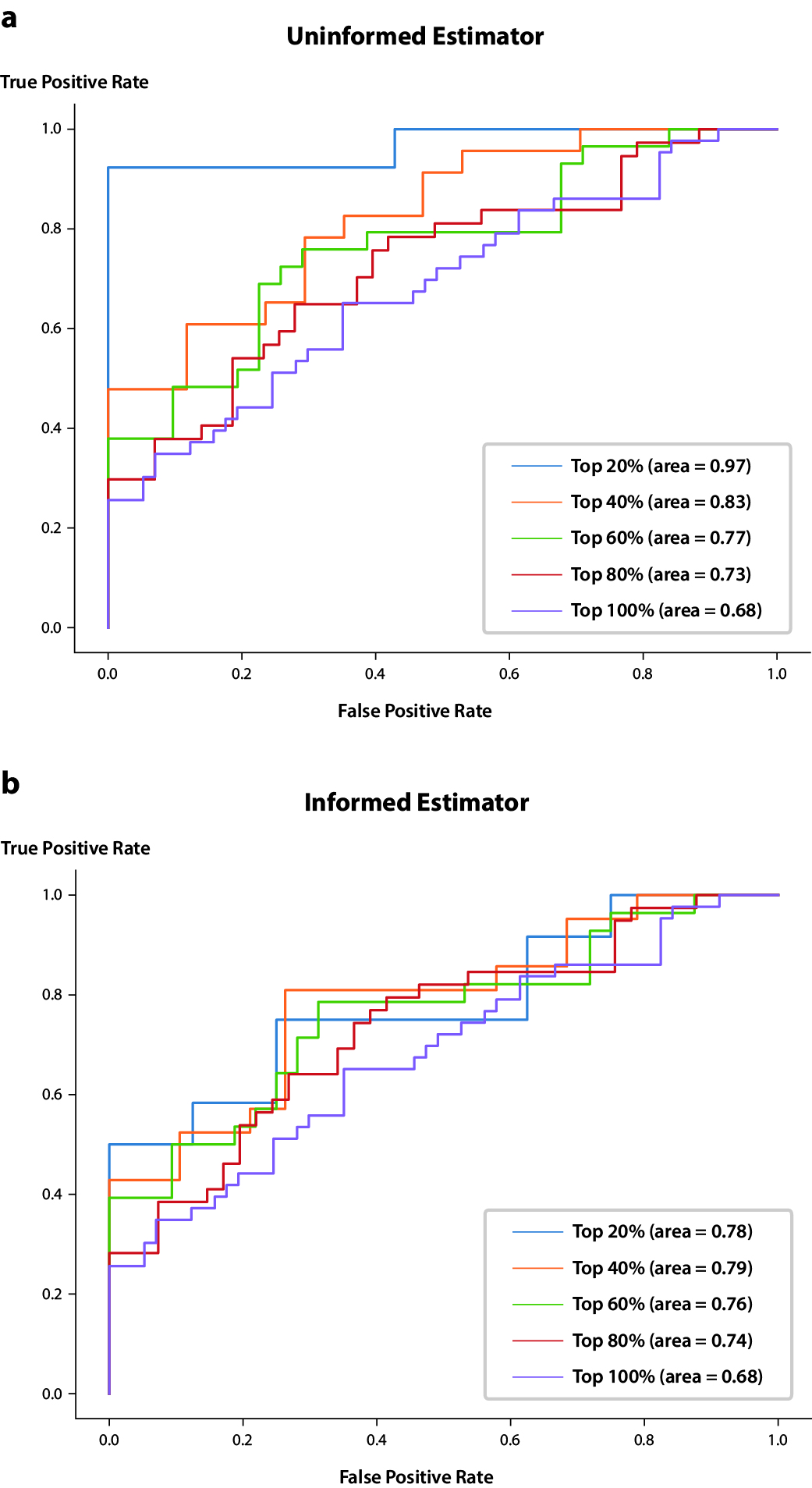}
  \caption{\textbf{Stratified ROC curves according to the retaining fraction of dataset.} Based on the sorted samples according to the ascending uncertainty scores, ROC curves of each subgroup that retained the top 20\%, 40\%, 60\%, 80\%, and 100\% (all samples) of the testing dataset are plotted. For example, the top 20\% subgroup consists of the first 20\% of samples with the smallest uncertainty scores. Uncertainty score from \textbf{(a)} the uninformed estimator and from \textbf{(b)} the informed estimator are utilized. As noted, the uncertainty scores from the uninformed estimator provided more informative stratification, where the ROC-AUC values progressively decreased from $0.97$ (top 20\%) to $0.68$ (top 100\%) as the cutoff of the uncertainty score increased. In contrast, the ROC curves based on the uncertainty scores from the uninformed estimator overlapped with each other, without providing such large progressive relationships. Therefore, the anticipated beneficial effect arising from the additional information related to the agreement between model explainability and physician understanding was not evident.}
  \label{fig:stratified_roc}
\end{figure}

\subsection{Stratified Predictive Performance by Uncertainty Scores}\label{sec:stratified_analysis}

Finally, we conducted an evaluation according to the cutoff of each acceptable uncertainty score for the predictive performance of the multimodal transformer on pelvic lymph node metastasis. We calculated the ROC-AUC for each subgroup that retained the top 20\%, 40\%, 60\%, 80\%, and 100\% (all samples) of samples in the testing dataset sorted in ascending order of uncertainty score. For example, the top 20\% subgroup consists of the first 20\% of samples with the smallest uncertainty scores. The uncertainty scores derived from both the uninformed estimator and the informed estimator were utilized in \textbf{\cref{fig:stratified_roc}a} and \textbf{\cref{fig:stratified_roc}b}, respectively. As a result, with the uncertainty scores obtained from the uninformed estimator $\mathcal{U}(\sigma_\mathrm{prob})$, it was observed that the ROC-AUC values progressively decreased from $0.97$ (top 20\%) to $0.68$ (top 100\%) as the cutoff of the uncertainty score increased (i.e., by accepting higher uncertainty and retaining more data). On the other hand, with the uncertainty scores obtained from the informed estimator $\mathcal{U}(\sigma_\mathrm{prob}, \alpha_\mathrm{image}, \alpha_\mathrm{table})$, such progressive changes were not prominent according to the cutoff of the uncertainty score, and the ROC-AUC values were distributed within a relatively narrow range (from $0.68$ in top 100\% to $0.79$ in top 40\%). These results also confirmed that the additional benefit brought by the agreement between model explainability and physician understanding for uncertainty scores is not significant.

\section{Discussion}\label{sec:discussion}

In order to develop a safe and reliable AI system, a good explanation must not only be convincing to physicians but also be grounded in the premise that it is actually linked with correct predictions. We examined the attention maps visualized from a multimodal transformer as a mechanism of explainability, assessing whether they remain as inappropriate explanations (Doctor-Agreement Explanation, High Uncertainty Prediction) as shown in \textbf{\cref{fig:framework}b}, or could become appropriate explanations (Doctor-Agreement Explanation, Low Uncertainty Prediction) as shown in \textbf{\cref{fig:framework}c}, based on the proposed framework (see \textbf{\cref{fig:question}}).

The case study employed a multimodal transformer, which was trained to predict pathological pelvic lymph node metastasis using tabular clinical data and 3D MRI data (see \textbf{\cref{fig:architecture}}). The physician-centered evaluation revealed various levels of agreement between the visualized attention maps and physician understanding (see \textbf{\cref{fig:image_attention}} and \textbf{\cref{fig:table_attention}}). By training uncertainty estimators that can predict the uncertainty score, which is the normalized form of model uncertainty, we investigated whether the additional information derived from the agreement contributes to reducing model uncertainty. If the anticipated benefit were indeed the case, the informed estimator $\mathcal{U}(\sigma_\mathrm{prob}, \alpha_\mathrm{image}, \alpha_\mathrm{table})$, which incorporates such additional information, could provide more informative uncertainty scores compared to the uninformed estimator $\mathcal{U}(\sigma_\mathrm{prob})$. However, we found no significant difference between the two estimators (see \textbf{\cref{fig:result_uncertainty_estimator}} and \textbf{\cref{fig:stratified_roc}}). Therefore, the additional information derived from the agreement did not significantly reduce the model's uncertainty, which might pose a risk for practitioners making decisions based on the superficially compelling model explanations, as illustrated in \textbf{\cref{fig:framework}b}. Given that evidence based on physician-centered evaluation of explainability is quite limited, we believe that sharing this study with the research community is crucial. It underscores the need for further improvement in explainability techniques that are truly beneficial for clinical decision-making.

\subsection{On the Prediction Performance of the Multimodal Transformer}

For the case study, we considered it more appropriate to select a relatively difficult task that includes hard samples with a wide range of uncertainty, rather than easier tasks. This is because the uncertainty associated with easy samples is expected to be limited, and in such cases, the impact of additional information on model uncertainty would likely exert less influence on the final model prediction, making it difficult to observe. Therefore, the task of predicting pathological pelvic lymph node metastasis from pre-operative information was selected.

Predicting pathological pelvic lymph node metastasis from preoperative information is generally a challenging task, owing to the limited information provided by MRI and clinical data, along with the occurrence of false positives and negatives \citep{Guo2012, Stijns2021}. In previous research, a deep-learning model trained with segmentation labels reported precision and recall values in the 0.6 range in an external testing dataset \citep{Zhao2020}. Indeed, the overall predictive performance of our multimodal transformer is at an intermediate level, as detailed in \textbf{\cref{sec:result_predictive_performance}}. Our result is based on an experimental setting that uses only classification labels, which have a coarser granularity compared to segmentation labels. Therefore, we consider that this level of performance may be deemed acceptable within the given setting.

Additionally, in a systematic review of human diagnostic performance using MRI for lymph node metastasis based on various morphological criteria, the summarized ROC curve had an AUC value around 0.79 \citep{Zhuang2021}. Compared to the results shown in \textbf{\cref{fig:stratified_roc}a}, our multimodal transformer exhibited the same level of performance for the top 60\% of the less uncertain samples, even without supervision based on the morphological criteria. Since achieving a high classification performance is not our primary goal, the intermediate level of classification performance was considered suitable for investigating the potential benefits brought about by the additional information regarding the agreement between model explainability and physician understanding.

\subsection{Importance of Guaranteeing Model Performance in Explainability}

While explainability in AI for healthcare has been emphasized as essential \citep{Kundu2021}, there is still no unified perspective on what attributes explainability should include. Notably, AI systems sometimes produce superficially convincing explanations that are not factually rooted in actual relationships \citep{Huang2023}. Therefore, as shown in \textbf{\cref{fig:question}}, we consider that reaching agreement between model explainability and physician understanding should provide additional information that reduces model uncertainty, thus ensuring safety when taking action based on the prediction.

This perspective is also justified from a cognitive science viewpoint. For example, automation bias is one of the most well-known sources of error in human-AI interaction, where humans tend to over-trust outputs from AI systems \citep{Ruskin2020}. In fact, the risk of automation bias has also been recognized in radiological diagnosis, where even experienced radiologists reading mammograms are prone to automation bias when supported by AI systems \citep{Dratsch2023}. Hence, explanations that are ``only superficially'' convincing can exacerbate the problem of automation bias, as such explanations would misleadingly enhance the acceptability of the AI system by human experts \citep{Lakkaraju2020} (see \textbf{\cref{fig:framework}b}). Therefore, when considering model explainability, it is essential to ensure that explanations convincing to humans are actually accompanied by safe predictions with low uncertainty (see \textbf{\cref{fig:framework}c}).

\subsection{Future Implications for Designing Good Explanations}

In this case study, we treated the state-of-the-art visualization technique of Transformer Explainability \citep{Chefer2021} as a method to visualize the attention map with gradient information. We believe that choosing this technique for model explainability is meaningful, as the majority of controversial evidence stems from saliency maps of CNNs \citep{Groen2022}, apart from the recently popularized transformers. However, since we observed no beneficial impact on reducing model uncertainty from the attention map, such a post hoc approach may face fundamental challenges in offering reliable explainability with performance guarantees. Indeed, several researchers have highlighted the numerous problems with many post hoc approaches and emphasized the importance of ante hoc approaches instead, which incorporate mechanisms of explainability by design \citep{Rudin2019, Kundu2021, Ghassemi2021}. We also consider that future research should focus on determining which attributes of explainability are desirable in clinical decision-making and on how to implement them by design.

\subsection{Limitations}

This study has several limitations. The case study is based on a dataset from a single facility, the deep-learning model is limited to a transformer-based architecture, and the technique employed for model explainability is restricted to one approach. Therefore, our observations might not apply to all post hoc approaches with various model architectures. The predictive performance of the multimodal transformer is moderate, and it is possible that the effect of explainability might differ in models with higher performance. However, it is very costly to have expert physicians evaluate the appropriateness of various variations of model explainability on a case-by-case basis. In fact, in our study, three experts evaluated a total of 160 MRI series, comprising 60 series from the validation dataset and 100 from the testing dataset (see \textbf{step 4} in \textbf{\cref{fig:dataset}}). Such exhaustive, physician-centered evaluations are very rare. We believe this study merits reporting to the research community due to its comprehensive evaluation of one of the state-of-the-art model explainability techniques from a clinical perspective.

\section{Conclusion}
Desirable model explainability that can be truly beneficial in clinical decision-making should lead to a reduction in model uncertainty by providing additional information regarding the agreement between model explainability and physician understanding. However, in our case study, we did not observe the anticipated beneficial effect. Such superficially compelling explanations could do more harm than good for patients by misleading physicians to rely on uncertain predictions. While explainability is essential in AI systems in medicine, the current state of the technology should not be overvalued, and there is still considerable room for future improvement.

\section*{Acknowledgement}
The authors thank the members of the Division of Medical AI Research and Development of the National Cancer Center Research Institute for their kind support.

\subsection*{Funding}
This work was supported by JST ACT-X (Grant Number JPMJAX23C9), JSPS KAKENHI (Grant Number JP22K07681), JST CREST (Grant Number JPMJCR1689), JST AIP-PRISM (Grant Number JPMJCR18Y4), and the National Cancer Center Research and Development Fund (2023-A-19).

\subsection*{Competing interests}
Kazuma Kobayashi and Ryuji Hamamoto have received research funding from Fujifilm Corporation.

\subsection*{Contributions}
K.K. conceived the study, devised the algorithms, coordinated the physician-centered evaluation, analyzed the results, and prepared the manuscript. Y.T., M.M., and S.I. prepared the datasets. K.K., Y.T., and M.M. joined the physician-centered evaluation as evaluators. L.G., T.N., and Y.A. provided technical advice. K.K. and T.N. implemented the algorithms and performed the technical evaluation. T.H., Y.K., and R.H. supervised the research.

\bibliography{reference}

\begin{thebibliography}{10}
\expandafter\ifx\csname url\endcsname\relax
  \def\url#1{\texttt{#1}}\fi
\expandafter\ifx\csname urlprefix\endcsname\relax\def\urlprefix{URL }\fi
\expandafter\ifx\csname href\endcsname\relax
  \def\href#1#2{#2} \def\path#1{#1}\fi

\bibitem{Kundu2021}
S.~Kundu, \href{https://doi.org/10.1038/s41591-021-01461-z}{{AI in medicine must be explainable}}, Nat. Med. 27 (2021) 1328.
\newblock \href {https://doi.org/10.1038/s41591-021-01461-z} {\path{doi:10.1038/s41591-021-01461-z}}.
\newline\urlprefix\url{https://doi.org/10.1038/s41591-021-01461-z}

\bibitem{Adadi2018}
A.~Adadi, M.~Berrada, {Peeking Inside the Black-Box: A Survey on Explainable Artificial Intelligence (XAI)}, IEEE Access 6 (2018) 52138--52160.
\newblock \href {https://doi.org/10.1109/ACCESS.2018.2870052} {\path{doi:10.1109/ACCESS.2018.2870052}}.

\bibitem{Cutillo2020}
C.~M. Cutillo, K.~R. Sharma, L.~Foschini, S.~Kundu, M.~Mackintosh, K.~D. Mandl, T.~Beck, E.~Collier, C.~Colvis, K.~Gersing, V.~Gordon, R.~Jensen, B.~Shabestari, N.~Southall, \href{https://pubmed.ncbi.nlm.nih.gov/32258429/}{{Machine intelligence in healthcare-perspectives on trustworthiness, explainability, usability, and transparency}}, NPJ Digit. Med. 3 (12 2020).
\newblock \href {https://doi.org/10.1038/S41746-020-0254-2} {\path{doi:10.1038/S41746-020-0254-2}}.
\newline\urlprefix\url{https://pubmed.ncbi.nlm.nih.gov/32258429/}

\bibitem{Khullar2022}
D.~Khullar, L.~P. Casalino, Y.~Qian, Y.~Lu, H.~M. Krumholz, S.~Aneja, \href{https://doi.org/10.1001/jamanetworkopen.2022.10309}{{Perspectives of Patients About Artificial Intelligence in Health Care}}, JAMA Netw. Open 5 (2022) e2210309--e2210309.
\newblock \href {https://doi.org/10.1001/jamanetworkopen.2022.10309} {\path{doi:10.1001/jamanetworkopen.2022.10309}}.
\newline\urlprefix\url{https://doi.org/10.1001/jamanetworkopen.2022.10309}

\bibitem{Kundu2023}
S.~Kundu, \href{https://doi.org/10.1038/s41562-023-01711-9}{{Measuring trustworthiness is crucial for medical AI tools}}, Nat. Hum. Behav. 7 (2023) 1812--1813.
\newblock \href {https://doi.org/10.1038/s41562-023-01711-9} {\path{doi:10.1038/s41562-023-01711-9}}.
\newline\urlprefix\url{https://doi.org/10.1038/s41562-023-01711-9}

\bibitem{Markus2021}
A.~F. Markus, J.~A. Kors, P.~R. Rijnbeek, \href{https://www.sciencedirect.com/science/article/pii/S1532046420302835}{{The role of explainability in creating trustworthy artificial intelligence for health care: A comprehensive survey of the terminology, design choices, and evaluation strategies}}, J. Biomed. Inform. 113 (2021) 103655.
\newblock \href {https://doi.org/10.1016/j.jbi.2020.103655} {\path{doi:10.1016/j.jbi.2020.103655}}.
\newline\urlprefix\url{https://www.sciencedirect.com/science/article/pii/S1532046420302835}

\bibitem{Ghassemi2021}
M.~Ghassemi, L.~Oakden-Rayner, A.~L. Beam, \href{https://www.sciencedirect.com/science/article/pii/S2589750021002089}{{The false hope of current approaches to explainable artificial intelligence in health care}}, Lancet Digit. Health 3 (2021) e745--e750.
\newblock \href {https://doi.org/10.1016/S2589-7500(21)00208-9} {\path{doi:10.1016/S2589-7500(21)00208-9}}.
\newline\urlprefix\url{https://www.sciencedirect.com/science/article/pii/S2589750021002089}

\bibitem{Rudin2019}
C.~Rudin, \href{https://doi.org/10.1038/s42256-019-0048-x}{{Stop explaining black box machine learning models for high stakes decisions and use interpretable models instead}}, Nat. Mach. Intell. 1 (2019) 206--215.
\newblock \href {https://doi.org/10.1038/s42256-019-0048-x} {\path{doi:10.1038/s42256-019-0048-x}}.
\newline\urlprefix\url{https://doi.org/10.1038/s42256-019-0048-x}

\bibitem{Groen2022}
A.~M. Groen, R.~Kraan, S.~F. Amirkhan, J.~G. Daams, M.~Maas, \href{https://pubmed.ncbi.nlm.nih.gov/36371947/}{{A systematic review on the use of explainability in deep learning systems for computer aided diagnosis in radiology: Limited use of explainable AI?}}, Eur. J. Radiol. 157 (12 2022).
\newblock \href {https://doi.org/10.1016/J.EJRAD.2022.110592} {\path{doi:10.1016/J.EJRAD.2022.110592}}.
\newline\urlprefix\url{https://pubmed.ncbi.nlm.nih.gov/36371947/}

\bibitem{Zhou2021}
J.~Zhou, A.~H. Gandomi, F.~Chen, A.~Holzinger, \href{https://www.mdpi.com/2079-9292/10/5/593}{{Evaluating the Quality of Machine Learning Explanations: A Survey on Methods and Metrics}}, Electronics 10~(5) (2021).
\newblock \href {https://doi.org/10.3390/electronics10050593} {\path{doi:10.3390/electronics10050593}}.
\newline\urlprefix\url{https://www.mdpi.com/2079-9292/10/5/593}

\bibitem{Huang2023}
L.~Huang, W.~Yu, W.~Ma, W.~Zhong, Z.~Feng, H.~Wang, Q.~Chen, W.~Peng, X.~Feng, B.~Qin, T.~Liu, {A Survey on Hallucination in Large Language Models: Principles, Taxonomy, Challenges, and Open Questions} (2023).
\newblock \href {http://arxiv.org/abs/2311.05232} {\path{arXiv:2311.05232}}, \href {https://doi.org/10.48550/arXiv.2311.05232} {\path{doi:10.48550/arXiv.2311.05232}}.

\bibitem{Lakkaraju2020}
H.~Lakkaraju, O.~Bastani, \href{https://doi.org/10.1145/3375627.3375833}{{"How Do I Fool You?": Manipulating User Trust via Misleading Black Box Explanations}}, in: Proceedings of the AAAI/ACM Conference on AI, Ethics, and Society (AIES), 2020, p. 79–85.
\newblock \href {https://doi.org/10.1145/3375627.3375833} {\path{doi:10.1145/3375627.3375833}}.
\newline\urlprefix\url{https://doi.org/10.1145/3375627.3375833}

\bibitem{Loftus2022}
T.~J. Loftus, B.~Shickel, M.~M. Ruppert, J.~A. Balch, T.~Ozrazgat-Baslanti, P.~J. Tighe, P.~A. Efron, W.~R. Hogan, P.~Rashidi, J.~G.~R. Upchurch, A.~Bihorac, \href{/pmc/articles/PMC9802673/ /pmc/articles/PMC9802673/?report=abstract https://www.ncbi.nlm.nih.gov/pmc/articles/PMC9802673/}{{Uncertainty-aware deep learning in healthcare: A scoping review}}, PLOS Digit. Health 1 (2022) e0000085.
\newblock \href {https://doi.org/10.1371/JOURNAL.PDIG.0000085} {\path{doi:10.1371/JOURNAL.PDIG.0000085}}.
\newline\urlprefix\url{/pmc/articles/PMC9802673/ /pmc/articles/PMC9802673/?report=abstract https://www.ncbi.nlm.nih.gov/pmc/articles/PMC9802673/}

\bibitem{Errit1998}
G.~E.~J. An, L.~Iefers, A.~N. M, A.~C. L.~J. Ansen, C.~Ornelis, J.~H. V. D.~V. Elde, J.~O.~H. Ermans, J.~Ohannes, H.~J. M. V.~K. Rieken, C.~Ees, J.~C. Ornelisse, R.~Ob, A.~E. M.~T. Ollenaar, \href{https://www.nejm.org/doi/full/10.1056/nejm199807233390403}{{Micrometastases and Survival in Stage II Colorectal Cancer}}, N. Engl. J. Med. 339 (1998) 223--228.
\newblock \href {https://doi.org/10.1056/NEJM199807233390403} {\path{doi:10.1056/NEJM199807233390403}}.
\newline\urlprefix\url{https://www.nejm.org/doi/full/10.1056/nejm199807233390403}

\bibitem{Zaborowski2021}
A.~M. Zaborowski, A.~Abdile, M.~Adamina, F.~Aigner, L.~D'Allens, C.~Allmer, A.~Álvarez, R.~Anula, M.~Andric, S.~Atallah, S.~Bach, M.~Bala, M.~Barussaud, A.~Bausys, B.~Bebington, A.~Beggs, F.~Bellolio, M.~R. Bennett, A.~Berdinskikh, V.~Bevan, S.~Biondo, G.~Bislenghi, M.~Bludau, A.~Boutall, N.~Brouwer, C.~Brown, C.~Bruns, D.~D. Buchanan, P.~Buchwald, J.~W. Burger, N.~Burlov, M.~Campanelli, M.~Capdepont, M.~Carvello, H.~H. Chew, D.~Christoforidis, D.~Clark, M.~Climent, K.~G. Cologne, T.~Contreras, R.~Croner, I.~R. Daniels, G.~Dapri, J.~Davies, P.~Delrio, Q.~Denost, M.~Deutsch, A.~Dias, A.~D'Hoore, E.~Drozdov, D.~Duek, M.~Dunlop, A.~Dziki, A.~Edmundson, S.~Efetov, A.~El-Hussuna, B.~Elliot, S.~Emile, E.~Espin, M.~Evans, S.~Faes, O.~Faiz, F.~Fleming, C.~Foppa, G.~Fowler, M.~Frasson, N.~Figueiredo, T.~Forgan, F.~Frizelle, S.~Gadaev, J.~Gellona, T.~Glyn, J.~Gong, B.~Goran, E.~Greenwood, M.~G. Guren, S.~Guillon, I.~Gutlic, D.~Hahnloser, H.~Hampel, A.~Hanly, H.~Hasegawa, L.~H. Iversen, A.~Hill, J.~Hill, J.~Hoch,
  M.~Hoffmeister, R.~Hompes, L.~Hurtado, F.~Iaquinandi, U.~Imbrasaite, R.~Islam, M.~D. Jafari, Y.~Kanemitsu, A.~Karachun, A.~A. Karimuddin, D.~S. Keller, J.~Kelly, R.~Kennelly, G.~Khrykov, P.~Kocian, C.~Koh, N.~Kok, K.~A. Knight, J.~Knol, C.~Kontovounisios, H.~Korner, Z.~Krivokapic, I.~Kronberger, H.~M. Kroon, M.~Kryzauskas, S.~Kural, M.~Kusters, Z.~Lakkis, T.~Lankov, D.~Larson, G.~Lázár, K.~Y. Lee, S.~H. Lee, J.~H. Lefèvre, A.~Lepisto, C.~Lieu, L.~Loi, C.~Lynch, H.~Maillou-Martinaud, A.~Maroli, S.~Martin, A.~Martling, K.~E. Matzel, J.~Mayol, F.~McDermott, G.~Meurette, M.~Millan, M.~Mitteregger, A.~Moiseenko, J.~R. Monson, S.~Morarasu, K.~Moritani, G.~Möslein, M.~Munini, C.~Nahas, S.~Nahas, I.~Negoi, A.~Novikova, M.~Ocares, K.~Okabayashi, A.~Olkina, L.~Oñate-Ocaña, J.~Otero, C.~Ozen, U.~Pace, G.~P.~S. Julião, L.~Panaiotti, Y.~Panis, D.~Papamichael, J.~Park, S.~Patel, J.~C.~P. Uriburu, M.~Pera, R.~O. Perez, A.~Petrov, F.~Pfeffer, P.~T. Phang, T.~Poskus, H.~Pringle, D.~Proud, I.~Raguz, N.~Rama,
  S.~Rasheed, M.~J. Raval, D.~Rega, C.~Reissfelder, J.~C.~R. Meneses, F.~Ris, S.~Riss, H.~Rodriguez-Zentner, C.~S. Roxburgh, A.~Saklani, A.~J. Salido, T.~Sammour, D.~Saraste, M.~Schneider, R.~Seishima, A.~Sekulic, T.~Seppala, K.~Sheahan, R.~Shine, A.~Shlomina, G.~S. Sica, T.~Singnomklao, L.~Siragusa, N.~Smart, A.~Solis, A.~Spinelli, R.~D. Staiger, M.~J. Stamos, S.~Steele, M.~Sunderland, K.~K. Tan, P.~J. Tanis, P.~Tekkis, B.~Teklay, S.~Tengku, M.~Jiménez-Toscano, P.~Tsarkov, M.~Turina, A.~Ulrich, B.~B. Vailati, M.~V. Harten, C.~Verhoef, S.~Warrier, S.~Wexner, H.~D. Wilt, B.~A. Weinberg, C.~Wells, A.~Wolthuis, E.~Xynos, N.~You, A.~Zakharenko, J.~Zeballos, D.~C. Winter, \href{https://jamanetwork.com/journals/jamasurgery/fullarticle/2781485}{{Characteristics of Early-Onset vs Late-Onset Colorectal Cancer: A Review}}, JAMA Surg. 156 (2021) 865--874.
\newblock \href {https://doi.org/10.1001/JAMASURG.2021.2380} {\path{doi:10.1001/JAMASURG.2021.2380}}.
\newline\urlprefix\url{https://jamanetwork.com/journals/jamasurgery/fullarticle/2781485}

\bibitem{Malla2023}
M.~Malla, K.~S. Pedersen, A.~R. Parikh, \href{https://jnccn.org/view/journals/jnccn/21/5.5/article-p567.xml}{{Updates in the Treatment of Metastatic Colorectal Cancer}}, J. Natl. Compr. Canc. Netw. 21 (2023) 567--571.
\newblock \href {https://doi.org/10.6004/JNCCN.2023.5012} {\path{doi:10.6004/JNCCN.2023.5012}}.
\newline\urlprefix\url{https://jnccn.org/view/journals/jnccn/21/5.5/article-p567.xml}

\bibitem{Guo2012}
X.~Guo, C.~Wang, X.~G. Shen, S.~Q. Ding, Y.~Y. Yu, Z.~G. Zhou, \href{/pmc/articles/PMC3362382/ /pmc/articles/PMC3362382/?report=abstract https://www.ncbi.nlm.nih.gov/pmc/articles/PMC3362382/}{{Occult tumor metastasis and the prognostic value of sentinel lymph nodes in rectal cancer}}, Oncol. Lett. 3 (2012) 411.
\newblock \href {https://doi.org/10.3892/OL.2011.490} {\path{doi:10.3892/OL.2011.490}}.
\newline\urlprefix\url{/pmc/articles/PMC3362382/ /pmc/articles/PMC3362382/?report=abstract https://www.ncbi.nlm.nih.gov/pmc/articles/PMC3362382/}

\bibitem{Stijns2021}
R.~C. Stijns, B.~W. Philips, I.~D. Nagtegaal, F.~Polat, J.~H. de~Wilt, C.~A. Wauters, P.~Zamecnik, J.~J. Fütterer, T.~W. Scheenen, {USPIO-enhanced MRI of lymph nodes in rectal cancer: A node-to-node comparison with histopathology}, Eur. J. Radiol. 138 (2021) 109636.
\newblock \href {https://doi.org/10.1016/J.EJRAD.2021.109636} {\path{doi:10.1016/J.EJRAD.2021.109636}}.

\bibitem{Vaswani2017}
A.~Vaswani, N.~Shazeer, N.~Parmar, J.~Uszkoreit, L.~Jones, A.~N. Gomez, Łukasz Kaiser, I.~Polosukhin, \href{https://proceedings.neurips.cc/paper_files/paper/2017/file/3f5ee243547dee91fbd053c1c4a845aa-Paper.pdf}{{Attention is All you Need}}, in: Advances in Neural Information Processing Systems (NIPS), Vol.~30, 2017.
\newline\urlprefix\url{https://proceedings.neurips.cc/paper_files/paper/2017/file/3f5ee243547dee91fbd053c1c4a845aa-Paper.pdf}

\bibitem{Devlin2019}
J.~Devlin, M.-W. Chang, K.~Lee, K.~Toutanova, \href{https://aclanthology.org/N19-1423}{{BERT: Pre-training of Deep Bidirectional Transformers for Language Understanding}}, in: North American Chapter of the Association for Computational Linguistics (NAACL), 2019, pp. 4171--4186.
\newblock \href {https://doi.org/10.18653/v1/N19-1423} {\path{doi:10.18653/v1/N19-1423}}.
\newline\urlprefix\url{https://aclanthology.org/N19-1423}

\bibitem{Radford2018}
A.~Radford, K.~Narasimhan, \href{https://s3-us-west-2.amazonaws.com/openai-assets/research-covers/language-unsupervised/language_understanding_paper.pdf}{{Improving Language Understanding by Generative Pre-Training}}, Open AI, 2018.
\newline\urlprefix\url{https://s3-us-west-2.amazonaws.com/openai-assets/research-covers/language-unsupervised/language_understanding_paper.pdf}

\bibitem{Xu2023}
P.~Xu, X.~Zhu, D.~A. Clifton, \href{https://doi.org/10.1109/TPAMI.2023.3275156,}{{Multimodal Learning With Transformers: A Survey}}, IEEE Trans. Pattern Anal. Mach. Intell. 45 (2023) 12113--12132.
\newblock \href {https://doi.org/10.1109/TPAMI.2023.3275156} {\path{doi:10.1109/TPAMI.2023.3275156}}.
\newline\urlprefix\url{https://doi.org/10.1109/TPAMI.2023.3275156,}

\bibitem{Alexey2021}
A.~Dosovitskiy, L.~Beyer, A.~Kolesnikov, D.~Weissenborn, X.~Zhai, T.~Unterthiner, M.~Dehghani, M.~Minderer, G.~Heigold, S.~Gelly, J.~Uszkoreit, N.~Houlsby, \href{https://openreview.net/forum?id=YicbFdNTTy}{{An Image is Worth 16x16 Words: Transformers for Image Recognition at Scale}}, in: International Conference on Learning Representations (ICLR), 2021.
\newline\urlprefix\url{https://openreview.net/forum?id=YicbFdNTTy}

\bibitem{Zhou2023}
H.-Y. Zhou, Y.~Yu, C.~Wang, S.~Zhang, Y.~Gao, J.~Pan, J.~Shao, G.~Lu, K.~Zhang, W.~Li, \href{https://doi.org/10.1038/s41551-023-01045-x}{{A transformer-based representation-learning model with unified processing of multimodal input for clinical diagnostics}}, Nat. Biomed. Eng. 7 (2023) 743--755.
\newblock \href {https://doi.org/10.1038/s41551-023-01045-x} {\path{doi:10.1038/s41551-023-01045-x}}.
\newline\urlprefix\url{https://doi.org/10.1038/s41551-023-01045-x}

\bibitem{Wang2019}
G.~Wang, W.~Li, M.~Aertsen, J.~Deprest, S.~Ourselin, T.~Vercauteren, \href{https://www.sciencedirect.com/science/article/pii/S0925231219301961}{{Aleatoric uncertainty estimation with test-time augmentation for medical image segmentation with convolutional neural networks}}, Neurocomputing 338 (2019) 34--45.
\newblock \href {https://doi.org/10.1016/j.neucom.2019.01.103} {\path{doi:10.1016/j.neucom.2019.01.103}}.
\newline\urlprefix\url{https://www.sciencedirect.com/science/article/pii/S0925231219301961}

\bibitem{Chefer2021}
H.~Chefer, S.~Gur, L.~Wolf, {Transformer Interpretability Beyond Attention Visualization}, in: IEEE/CVF Conference on Computer Vision and Pattern Recognition (CVPR), 2021, pp. 782--791.
\newblock \href {https://doi.org/10.1109/CVPR46437.2021.00084} {\path{doi:10.1109/CVPR46437.2021.00084}}.

\bibitem{Patr2023}
C.~Patr\'{\i}cio, J.~a.~C. Neves, L.~F. Teixeira, \href{https://doi.org/10.1145/3625287}{{Explainable Deep Learning Methods in Medical Image Classification: A Survey}}, ACM Comput. Surv. 56~(4) (oct 2023).
\newblock \href {https://doi.org/10.1145/3625287} {\path{doi:10.1145/3625287}}.
\newline\urlprefix\url{https://doi.org/10.1145/3625287}

\bibitem{Kim2015}
B.~Kim, C.~Rudin, J.~Shah, \href{https://dl.acm.org/doi/10.5555/2969033.2969045}{{The Bayesian Case Model: A Generative Approach for Case-Based Reasoning and Prototype Classification}}, Advances in Neural Information Processing Systems (NIPS) 3 (3 2015).
\newline\urlprefix\url{https://dl.acm.org/doi/10.5555/2969033.2969045}

\bibitem{Caruana2015}
R.~Caruana, Y.~Lou, J.~Gehrke, P.~Koch, M.~Sturm, N.~Elhadad, \href{https://doi.org/10.1145/2783258.2788613}{{Intelligible Models for HealthCare: Predicting Pneumonia Risk and Hospital 30-Day Readmission}}, in: ACM SIGKDD International Conference on Knowledge Discovery and Data Mining (KDD), 2015, pp. 1721--1730.
\newblock \href {https://doi.org/10.1145/2783258.2788613} {\path{doi:10.1145/2783258.2788613}}.
\newline\urlprefix\url{https://doi.org/10.1145/2783258.2788613}

\bibitem{Krishna2022}
S.~Krishna, T.~Han, A.~Gu, J.~Pombra, S.~Jabbari, S.~Wu, H.~Lakkaraju, {The Disagreement Problem in Explainable Machine Learning: A Practitioner's Perspective} (2022).
\newblock \href {http://arxiv.org/abs/2202.01602} {\path{arXiv:2202.01602}}, \href {https://doi.org/10.48550/arXiv.2202.01602} {\path{doi:10.48550/arXiv.2202.01602}}.

\bibitem{Gilpin2019}
L.~H. Gilpin, D.~Bau, B.~Z. Yuan, A.~Bajwa, M.~Specter, L.~Kagal, {Explaining Explanations: An Overview of Interpretability of Machine Learning}, in: International Conference on Data Science and Advanced Analytics (DSAA), 2018, pp. 80--89.
\newblock \href {https://doi.org/10.1109/DSAA.2018.00018} {\path{doi:10.1109/DSAA.2018.00018}}.

\bibitem{Zana2020}
Z.~Bu\c{c}inca, P.~Lin, K.~Z. Gajos, E.~L. Glassman, \href{https://doi.org/10.1145/3377325.3377498}{{Proxy Tasks and Subjective Measures Can Be Misleading in Evaluating Explainable AI Systems}}, in: International Conference on Intelligent User Interfaces (IUI), 2020, p. 454–464.
\newblock \href {https://doi.org/10.1145/3377325.3377498} {\path{doi:10.1145/3377325.3377498}}.
\newline\urlprefix\url{https://doi.org/10.1145/3377325.3377498}

\bibitem{Doshivelez2017}
F.~Doshi-Velez, B.~Kim, {Towards A Rigorous Science of Interpretable Machine Learning} (2017).
\newblock \href {http://arxiv.org/abs/1702.08608} {\path{arXiv:1702.08608}}, \href {https://doi.org/10.48550/arXiv.1702.08608} {\path{doi:10.48550/arXiv.1702.08608}}.

\bibitem{Tonekaboni2019}
S.~Tonekaboni, S.~Joshi, M.~Mccradden, A.~Goldenberg, \href{https://proceedings.mlr.press/v106/tonekaboni19a/tonekaboni19a.pdf}{{What Clinicians Want: Contextualizing Explainable Machine Learning for Clinical End Use}}, in: Machine Learning for Healthcare (MLHC), 2019.
\newline\urlprefix\url{https://proceedings.mlr.press/v106/tonekaboni19a/tonekaboni19a.pdf}

\bibitem{Dong2023}
Z.~Dong, J.~Wang, Y.~Li, Y.~Deng, W.~Zhou, X.~Zeng, D.~Gong, J.~Liu, J.~Pan, R.~Shang, Y.~Xu, M.~Xu, L.~Zhang, M.~Zhang, X.~Tao, Y.~Zhu, H.~Du, Z.~Lu, L.~Yao, L.~Wu, H.~Yu, \href{https://doi.org/10.1038/s41746-023-00813-y}{{Explainable artificial intelligence incorporated with domain knowledge diagnosing early gastric neoplasms under white light endoscopy}}, Npj Digit. Med. 6 (2023) 64.
\newblock \href {https://doi.org/10.1038/s41746-023-00813-y} {\path{doi:10.1038/s41746-023-00813-y}}.
\newline\urlprefix\url{https://doi.org/10.1038/s41746-023-00813-y}

\bibitem{Gaube2023}
S.~Gaube, H.~Suresh, M.~Raue, E.~Lermer, T.~K. Koch, M.~F.~C. Hudecek, A.~D. Ackery, S.~C. Grover, J.~F. Coughlin, D.~Frey, F.~C. Kitamura, M.~Ghassemi, E.~Colak, \href{https://doi.org/10.1038/s41598-023-28633-w}{{Non-task expert physicians benefit from correct explainable AI advice when reviewing X-rays}}, Sci. Rep. 13 (2023) 1383.
\newblock \href {https://doi.org/10.1038/s41598-023-28633-w} {\path{doi:10.1038/s41598-023-28633-w}}.
\newline\urlprefix\url{https://doi.org/10.1038/s41598-023-28633-w}

\bibitem{Tsiknakis2020}
N.~Tsiknakis, E.~Trivizakis, E.~E. Vassalou, G.~Z. Papadakis, D.~A. Spandidos, A.~Tsatsakis, J.~Sánchez-García, R.~López-González, N.~Papanikolaou, A.~H. Karantanas, K.~Marias, \href{/pmc/articles/PMC7388253/ /pmc/articles/PMC7388253/?report=abstract https://www.ncbi.nlm.nih.gov/pmc/articles/PMC7388253/}{{Interpretable artificial intelligence framework for COVID-19 screening on chest X-rays}}, Exp. Ther. Med. 20 (2020) 727.
\newblock \href {https://doi.org/10.3892/ETM.2020.8797} {\path{doi:10.3892/ETM.2020.8797}}.
\newline\urlprefix\url{/pmc/articles/PMC7388253/ /pmc/articles/PMC7388253/?report=abstract https://www.ncbi.nlm.nih.gov/pmc/articles/PMC7388253/}

\bibitem{Selvaraju2017}
R.~R. Selvaraju, M.~Cogswell, A.~Das, R.~Vedantam, D.~Parikh, D.~Batra, {Grad-CAM: Visual Explanations from Deep Networks via Gradient-Based Localization}, in: International Conference on Computer Vision (ICCV), 2017, pp. 618--626.
\newblock \href {https://doi.org/10.1109/ICCV.2017.74} {\path{doi:10.1109/ICCV.2017.74}}.

\bibitem{Ehrmann2023}
D.~E. Ehrmann, S.~Joshi, S.~D. Goodfellow, M.~L. Mazwi, D.~Eytan, \href{https://doi.org/10.1038/s41746-023-00753-7}{{Making machine learning matter to clinicians: model actionability in medical decision-making}}, Npj Digit. Med. 6 (2023) 7.
\newblock \href {https://doi.org/10.1038/s41746-023-00753-7} {\path{doi:10.1038/s41746-023-00753-7}}.
\newline\urlprefix\url{https://doi.org/10.1038/s41746-023-00753-7}

\bibitem{Jain2019}
S.~Jain, B.~C. Wallace, \href{https://aclanthology.org/N19-1357}{{Attention is not Explanation}}, in: North American Chapter of the Association for Computational Linguistics (NAACL), 2019, pp. 3543--3556.
\newblock \href {https://doi.org/10.18653/v1/N19-1357} {\path{doi:10.18653/v1/N19-1357}}.
\newline\urlprefix\url{https://aclanthology.org/N19-1357}

\bibitem{Wiegreffe2019}
S.~Wiegreffe, Y.~Pinter, \href{https://aclanthology.org/D19-1002}{{Attention is not not Explanation}}, in: Empirical Methods in Natural Language Processing and International Joint Conference on Natural Language Processing (EMNLP-IJCNLP), 2019, pp. 11--20.
\newblock \href {https://doi.org/10.18653/v1/D19-1002} {\path{doi:10.18653/v1/D19-1002}}.
\newline\urlprefix\url{https://aclanthology.org/D19-1002}

\bibitem{Zou2023}
K.~Zou, Z.~Chen, X.~Yuan, X.~Shen, M.~Wang, H.~Fu, \href{https://www.sciencedirect.com/science/article/pii/S2950162823000036}{{A review of uncertainty estimation and its application in medical imaging}}, Meta-Radiol. 1 (2023) 100003.
\newblock \href {https://doi.org/10.1016/j.metrad.2023.100003} {\path{doi:10.1016/j.metrad.2023.100003}}.
\newline\urlprefix\url{https://www.sciencedirect.com/science/article/pii/S2950162823000036}

\bibitem{Ma2022}
M.~Ma, J.~Ren, L.~Zhao, D.~Testuggine, X.~Peng, {Are Multimodal Transformers Robust to Missing Modality?}, in: IEEE/CVF Conference on Computer Vision and Pattern Recognition (CVPR), 2022, pp. 18156--18165.
\newblock \href {https://doi.org/10.1109/CVPR52688.2022.01764} {\path{doi:10.1109/CVPR52688.2022.01764}}.

\bibitem{Gal2016}
Y.~Gal, Z.~Ghahramani, \href{https://proceedings.mlr.press/v48/gal16.html}{{Dropout as a Bayesian Approximation: Representing Model Uncertainty in Deep Learning}}, in: International Conference on Machine Learning (ICML), Vol.~48 of Proceedings of Machine Learning Research, 2016, pp. 1050--1059.
\newline\urlprefix\url{https://proceedings.mlr.press/v48/gal16.html}

\bibitem{Gal2016_RNN}
Y.~Gal, Z.~Ghahramani, \href{https://dl.acm.org/doi/10.5555/3157096.3157211}{{A Theoretically Grounded Application of Dropout in Recurrent Neural Networks}}, in: Advances in Neural Information Processing Systems (NIPS), Vol.~29, 2016.
\newline\urlprefix\url{https://dl.acm.org/doi/10.5555/3157096.3157211}

\bibitem{Sankararaman2022}
K.~A. Sankararaman, S.~Wang, H.~Fang, {BayesFormer: Transformer with Uncertainty Estimation} (2022).
\newblock \href {http://arxiv.org/abs/2206.00826} {\path{arXiv:2206.00826}}, \href {https://doi.org/10.48550/arXiv.2206.00826} {\path{doi:10.48550/arXiv.2206.00826}}.

\bibitem{Barz2020}
B.~Barz, J.~Denzler, {Deep Learning on Small Datasets without Pre-Training using Cosine Loss}, in: Winter Conference on Applications of Computer Vision (WACV), 2020, pp. 1360--1369.
\newblock \href {https://doi.org/10.1109/WACV45572.2020.9093286} {\path{doi:10.1109/WACV45572.2020.9093286}}.

\bibitem{Lahoud20223}
J.~Lahoud, J.~Cao, F.~S. Khan, H.~Cholakkal, R.~M. Anwer, S.~Khan, M.-H. Yang, {3D Vision with Transformers: A Survey} (2022).
\newblock \href {http://arxiv.org/abs/2208.04309} {\path{arXiv:2208.04309}}, \href {https://doi.org/10.48550/arXiv.2208.04309} {\path{doi:10.48550/arXiv.2208.04309}}.

\bibitem{Gorishniy2023}
Y.~Gorishniy, I.~Rubachev, V.~Khrulkov, A.~Babenko, \href{https://proceedings.neurips.cc/paper_files/paper/2021/file/9d86d83f925f2149e9edb0ac3b49229c-Paper.pdf}{Revisiting deep learning models for tabular data}, in: Advances in Neural Information Processing Systems (NeurIPS), Vol.~34, 2021, pp. 18932--18943.
\newline\urlprefix\url{https://proceedings.neurips.cc/paper_files/paper/2021/file/9d86d83f925f2149e9edb0ac3b49229c-Paper.pdf}

\bibitem{Balntas2016}
V.~Balntas, E.~Riba, D.~Ponsa, K.~Mikolajczyk, {Learning local feature descriptors with triplets and shallow convolutional neural networks}, in: British Machine Vision Association (BMVA), 2016, pp. 119.1--119.11.
\newblock \href {https://doi.org/10.5244/C.30.119} {\path{doi:10.5244/C.30.119}}.

\bibitem{Kingma2017}
D.~P. Kingma, J.~Ba, {Adam: A Method for Stochastic Optimization}, in: International Conference on Learning Representations (ICLR), 2015.

\bibitem{Müller2020}
R.~M\"{u}ller, S.~Kornblith, G.~Hinton, \href{https://dl.acm.org/doi/10.5555/3454287.3454709}{{When Does Label Smoothing Help?}}, in: Advances in Neural Information Processing Systems (NeurIPS), 2019, pp. 4694--4703.
\newline\urlprefix\url{https://dl.acm.org/doi/10.5555/3454287.3454709}

\bibitem{Zhao2020}
X.~Zhao, P.~Xie, M.~Wang, W.~Li, P.~J. Pickhardt, W.~Xia, F.~Xiong, R.~Zhang, Y.~Xie, J.~Jian, H.~Bai, C.~Ni, J.~Gu, T.~Yu, Y.~Tang, X.~Gao, X.~Meng, \href{https://pubmed.ncbi.nlm.nih.gov/32512507/}{{Deep learning-based fully automated detection and segmentation of lymph nodes on multiparametric-mri for rectal cancer: A multicentre study}}, EBioMedicine 56 (6 2020).
\newblock \href {https://doi.org/10.1016/J.EBIOM.2020.102780} {\path{doi:10.1016/J.EBIOM.2020.102780}}.
\newline\urlprefix\url{https://pubmed.ncbi.nlm.nih.gov/32512507/}

\bibitem{Zhuang2021}
Z.~Zhuang, Y.~Zhang, M.~Wei, X.~Yang, Z.~Wang, \href{/pmc/articles/PMC8315047/ /pmc/articles/PMC8315047/?report=abstract https://www.ncbi.nlm.nih.gov/pmc/articles/PMC8315047/}{{Magnetic Resonance Imaging Evaluation of the Accuracy of Various Lymph Node Staging Criteria in Rectal Cancer: A Systematic Review and Meta-Analysis}}, Front. Oncol. 11 (2021) 709070.
\newblock \href {https://doi.org/10.3389/FONC.2021.709070} {\path{doi:10.3389/FONC.2021.709070}}.
\newline\urlprefix\url{/pmc/articles/PMC8315047/ /pmc/articles/PMC8315047/?report=abstract https://www.ncbi.nlm.nih.gov/pmc/articles/PMC8315047/}

\bibitem{Ruskin2020}
K.~J. Ruskin, C.~Corvin, S.~C. Rice, S.~R. Winter, {Autopilots in the Operating Room}, Anesthesiology 133 (2020) 653--665.
\newblock \href {https://doi.org/10.1097/ALN.0000000000003385} {\path{doi:10.1097/ALN.0000000000003385}}.

\bibitem{Dratsch2023}
T.~Dratsch, X.~Chen, M.~R. Mehrizi, R.~Kloeckner, A.~Mähringer-Kunz, M.~Püsken, B.~Baeßler, S.~Sauer, D.~Maintz, D.~P. dos Santos, {Automation Bias in Mammography: The Impact of Artificial Intelligence BI-RADS Suggestions on Reader Performance}, Radiology 307 (5 2023).
\newblock \href {https://doi.org/10.1148/radiol.222176} {\path{doi:10.1148/radiol.222176}}.

\end{thebibliography}

\end{document}